\documentclass[twocolumn,showkeys,showpacs,preprintnumbers,prd,superscriptaddress,nofootinbib,notitlepage]{revtex4-2}
\usepackage{graphicx,epsf,bm,amsmath,amsfonts,amssymb,epstopdf,natbib,hyperref,color,verbatim,multirow,bm,orcidlink}
\hypersetup{colorlinks=true,urlcolor=blue,citecolor=blue,linkcolor=blue,menucolor=blue,anchorcolor=blue,filecolor=blue}

\newcommand{\be}{\begin{equation}}
\newcommand{\ee}{\end{equation}}
\newcommand{\bs}{\begin{split}} 
\newcommand{\bea}{\begin{eqnarray}}
\newcommand{\eea}{\end{eqnarray}} 
\newcommand{\al}{\alpha}

\begin{document}

\title{Testing $\alpha$-attractor quintessential inflation against CMB and low-redshift data}

\author{William Giar\`{e}\orcidlink{0000-0002-4012-9285}}
\email{w.giare@sheffield.ac.uk}
\affiliation{School of Mathematics and Statistics, University of Sheffield, Hounsfield Road, Sheffield S3 7RH, United Kingdom \looseness=-1}

\author{Eleonora Di Valentino\orcidlink{0000-0001-8408-6961}}
\email{e.divalentino@sheffield.ac.uk}
\affiliation{School of Mathematics and Statistics, University of Sheffield, Hounsfield Road, Sheffield S3 7RH, United Kingdom \looseness=-1}

\author{Eric V.\ Linder\orcidlink{0000-0001-5536-9241}}
\email{evlinder@lbl.gov}
\affiliation{Berkeley Center for Cosmological Physics \& Berkeley Lab, University of California, Berkeley, CA 94720, USA}

\author{Enrico Specogna}
\email{especogna1@sheffield.ac.uk}
\affiliation{School of Mathematics and Statistics, University of Sheffield, Hounsfield Road, Sheffield S3 7RH, United Kingdom \looseness=-1}

\date{\today}

\begin{abstract}
\noindent Due to universality and attractor properties, $\alpha$-attractor quintessential inflation establishes direct relations between inflationary observables such as the scalar tilt $n_s$ and the tensor-to-scalar ratio $r$, and late-time dark energy equation of state parameters $w_0$ and $w_a$. In this work, we examine three different physically motivated regimes, considering complete freedom in the parameter $\alpha$, models inspired by supergravity where $\alpha$ takes on values up to $\alpha=7/3$, and Starobinsky inflation ($\alpha=1$). We investigate the consistency and constraints imposed by Cosmic Microwave Background measurements from the Planck satellite, B-mode polarization data from the BICEP/Keck collaboration, and low-redshift observations. Additionally, we consider small-scale CMB measurements released by the Atacama Cosmology Telescope, which give results approaching the Harrison-Zel'dovich spectrum ($n_s \approx 1$). Here $\alpha$-attractors lead to an improved fit over $\Lambda$CDM. For the large-scale CMB measurements,  $\alpha\gtrsim2$ models can provide equally good fits as $\Lambda$CDM.
\end{abstract}
\maketitle

\section{Introduction}
An undoubtedly fascinating aspect of modern cosmology is that, according to the standard $\Lambda$CDM model, our universe appears to have undergone two distinct phases of accelerated expansion.

The first one -- inflation~\cite{Guth:1980zm} -- is expected to occur at very early times and thus at very high energy scales. From a geometrical point of view, the inflationary universe was almost de Sitter (i.e., a maximally symmetric solution of the Einstein Equations with a positive cosmological constant). The rapid accelerated expansion experienced by the spacetime in this stage is expected to set the correct initial conditions for the subsequent Hot Big Bang theory evolution, stretching the universe towards spatial flatness and simultaneously solving other issues such as the horizon and entropy problems, and the apparent lack of topological defects \cite{Guth:1980zm, Linde:1981mu, Albrecht:1982wi}. On the other hand, at the perturbation level, inflation also offers an elegant mechanism to explain the physical origins of the first fluctuations in the universe, which eventually gave rise to observed structures such as galaxies and clusters of galaxies \cite{Mukhanov:1981xt, Bardeen:1983qw, Hawking:1982cz, Guth:1982ec}.

A second phase of accelerated expansion is instead observed to take place at present, as firstly determined via observations of distant Type Ia Supernovae \cite{SupernovaSearchTeam:1998fmf, SupernovaCosmologyProject:1998vns}, and subsequently (directly and indirectly) corroborated by a wide variety of other probes \cite{Sherwin:2011gv, Moresco:2016mzx, Haridasu:2017lma, Rubin:2016iqe, Planck:2018vyg, Gomez-Valent:2018gvm, Yang:2019fjt, Nadathur:2020kvq, Rose:2020shp, DiValentino:2020evt, eBOSS:2020yzd, Escamilla:2023oce}. 

The acceleration characterizing the dynamics of the universe, both in its early moments and at present, has garnered and continues to attract significant research interest within the cosmology and high-energy physics community. This interest is, in part, driven by the intrinsic implications of observations pointing towards an asymptotically de Sitter universe. This stands in contrast to expectations from extensions to fundamental physics, where several theories/models of quantum gravity typically propose an asymptotically anti-de Sitter universe. However, it is also motivated by the fact that all known forces and components in nature would decelerate the expansion of the Universe. Therefore, from a dynamical point of view, acceleration forces us to consider the existence of a physical mechanism beyond the standard model of fundamental interactions, able to induce a phase of repulsive gravity, whether through vacuum energy, new fields, or modified gravity. For instance, within the standard $\Lambda$CDM cosmological model, we assume that inflation is driven by a single scalar field minimally coupled to gravity, undergoing slow-roll evolution on its potential. Instead, to explain the second phase of accelerated expansion, the standard cosmological model introduces a form of Dark Energy (DE) represented by a positive cosmological constant $\Lambda$ in the Einstein equations.

Given our limited understanding of the microphysical realization of both epochs and the fact that, although profoundly differing in characteristic energy scales, Inflation and DE share several common aspects, an interesting possibility to explore is whether they could both originate from the same underlying physics. Although significant challenges arise when attempting to simultaneously explain two periods of accelerated expansion spanning energy scales that differ by over 100 orders of magnitude in characteristic energy density, several ideas have been proposed to connect them through particle physics or gravitation, see, e.g., Refs.~\cite{Peebles:1998qn,Baccigalupi:1998mn,Peloso:1999dm,Kaganovich:2000fc,Dimopoulos:2000md,Dimopoulos:2001ix,Nunes:2002wz,Nunes:2002wz,Dimopoulos:2002hm,Giovannini:2003jw,Tashiro:2003qp,Sami:2004xk,Rosenfeld:2005mt,BuenoSanchez:2006fhh,Neupane:2007mu,WaliHossain:2014usl,Hossain:2014coa,Dimopoulos:2017zvq,vandeBruck:2017voa,DeHaro:2017abf,Haro:2018zdb,Dimopoulos:2019gpz,Haro:2019gsv,Benisty:2020qta,Dimopoulos:2021xld,deHaro:2021swo,Bettoni:2021qfs,Das:2023nmm,Choudhury:2023kam}. Models in which DE can arise from the same potential that gave inflation at early times are typically referred to as quintessential inflation. In particular, $\alpha$-attractors ~\cite{Kallosh:2013hoa,Kallosh:2013yoa,Galante:2014ifa,Linder:2015qxa,Braglia:2020bym}
have been identified as a promising category of models to obtain quintessential inflation 
\cite{Dimopoulos:2017zvq,Dimopoulos:2017tud,Akrami:2017cir,Garcia-Garcia:2018hlc,Akrami:2020zxw,AresteSalo:2021wgb,Dimopoulos:2021xld,Zhumabek:2023wka}. Indeed, they can have desirable  properties such as robustness against quantum corrections,  shift symmetry, and attractor behavior enabling insensitivity to initial conditions. In addition, they can be highly predictive in their implications for both inflation and DE observables. This predictivity for instance implies that current measurement of the scalar power spectrum tilt $n_s$ will point toward allowed regions of the tensor to scalar power ratio $r$. In addition, since $\alpha$-attractor quintessential inflation is distinct from a cosmological constant, it predicts specific DE dynamics of a thawing field \cite{Akrami:2017cir,Zhumabek:2023wka}, with $w_0\ne-1$, $w_a\ne0$.  

As this model class is theoretically quite attractive, it is worthwhile to test it against observations. 
We confront $\alpha$-attractor quintessential inflation against the latest cosmic microwave background (CMB) and cosmic distance and redshift space distortion measurements. We aim to understand what insights these data can offer about the expected parameter range and whether upcoming experiments can detect them. Our approach involves a thorough and consistent analysis of the model, connecting different evolutionary phases as per theoretical predictions. To the best of our knowledge, this represents one of the first studies assuming quintessential inflation from the outset of the analysis, making it a pivotal work from which we can draw pioneering insights. The manuscript is organized as follows. In \autoref{sec:theory}, we review the most crucial theoretical predictions of $\alpha$-attractor quintessential inflation, along with its connection to inflation and DE quantities. Moving to \autoref{sec:method}, we explain how we consistently incorporate these relationships into the data analysis methodology. In \autoref{sec:planck}, we analyze observational constraints on the model based on Planck CMB measurements, considering various physically motivated choices for the prior range on the key $\alpha$ parameter. In \autoref{sec:ACT}, we repeat the analysis focusing instead on small-scale CMB measurements from the Atacama Cosmology Telescope (ACT) and examining the impact of higher $n_s$ typically favored by this dataset. Finally,~\autoref{sec:concl} summarizes the results and discusses implications for upcoming surveys.

\section{$\alpha$-Attractor Properties}  
\label{sec:theory} 

$\alpha$-attractors have a pole in their kinetic term, which 
when transformed to a canonical form involving 
$\phi=\sqrt{6\alpha}\,{\rm arctanh}(\varphi/\sqrt{6})$ stretches the canonical field value $\phi$ 
out to infinity, giving a long flat plateau ideal for inflation. 
The characteristic scale of the stretching is determined by 
the value $\alpha$ that can be identified with particular 
characteristics within particle physics theories such as 
supergravity \cite{Kallosh:2013hoa,Kallosh:2013yoa,Galante:2014ifa}. In particular, 
within maximal supergravity, $\alpha$ is related to Poincar{\'e} disks 
of the symmetry group and takes on discrete values 
$\alpha\in\{1/3\, , \, 2/3\, , \, \dots \,, \,7/3\}$. In this work, we will investigate results for three ranges of $\alpha$ to determine their impact on the observables: an uninformative prior $\alpha=[0,\alpha_{\rm max} \gg 1]$, a continuous range $\alpha=[0,7/3]$ inspired by supergravity, and a fixed value $\alpha=1$ corresponding to Starobinsky and Higgs inflation. 

Regardless of the value of $\alpha$, due to their pole structure $\alpha$-attractors have a universality class behavior 
\be 
n_s=1-\frac{2}{N}\ , 
\label{eq.ns}
\ee 
where $n_s$ is the scalar tilt of the primordial density power spectrum and $N$ is the number of e-folds of inflation between 
horizon crossing of the mode at the pivot scale $k_{\star}=0.05$ Mpc$^{-1}$
and the end of inflation. Furthermore, as the field rolls 
off the plateau to end inflation, the steepness involves 
$\alpha$, determining the tensor (primordial gravitational 
wave) to scalar (density) power 
\be 
r=\frac{12\alpha}{N^2} \ . 
\label{eq.r}
\ee 
The high Hubble friction during the radiation and matter 
dominated eras freeze the field and it is only released 
to roll (thaws) recently, putting it into the thawing class 
of DE dynamics \cite{Caldwell:2005tm,Linder:2006sv}. The pole 
structure does not determine the specific dynamics at late 
times, however: this depends on the form of the potential 
near $\phi=0$. Many arbitrary choices exist for this, but 
two desirable characteristics guide our choice. First, we 
do not allow an explicit cosmological constant -- that would  
exacerbate fine tuning and remove much of the motivation 
for any dynamics, making late time observations to be 
indistinguishable from a cosmological constant alone. 
Second, we seek some protection from quantum radiative 
corrections by employing a shift symmetry, and so we 
adopt a linear potential as initially used for inflation
\cite{Linde:1986dq} and explored for DE (e.g.\ 
\cite{Kallosh:2003bq,Kratochvil:2004gq}). Happily, after the field redefinition 
the linear potential becomes an exponential potential 
with attractor behavior. This is named the ExpLin 
$\al$-attractor \cite{Zhumabek:2023wka} 
(following \cite{Akrami:2017cir,Akrami:2020zxw}). 

The attractor enforces a DE equation of state 
$w(a\gg1)=-1+2/(9\alpha)$, creating a distinct offset from 
the cosmological constant value of $w=-1$. As demonstrated 
in \cite{Zhumabek:2023wka} from exact solutions of the Klein-Gordon 
field equation and Friedmann equations, 
as long as the field is frozen at large enough values  
the late time dynamics for viable models gives DE 
equation of state parameter relations 
\be 
w_0= -1+\frac{4}{3N^2\,r}\ , 
\label{eq.w0}
\ee 
and 
\be 
w_a\approx -1.53\,(1+w_0) \qquad \Longrightarrow \qquad w_a\approx-\frac{2}{N^2\,r}\ .
\label{eq.wa}
\ee 
Note that since the DE density contribution 
diminishes rapidly into the past, the exact form of $w(a)$ is 
unimportant for the CMB, just the distance to the last scattering 
surface. The $w_0$--$w_a$ parametrization has been demonstrated 
as accurate to 0.1\% for these purposes \cite{Linder:2002et,dePutter:2008wt}. 
Thus we see a close relation between the late time dark energy behavior and the early universe inflation characteristics, i.e.\ 
$(\alpha,N)\to (n_s,r)\to (w_0,w_a)$. 

\section{Methods and Data}
\label{sec:method}
To parametrize the effects of $\alpha$-attractor quintessential inflation on the inflationary observables, we assume the usual power-law spectrum of primordial adiabatic components, given by:
\begin{equation}
	\log \mathcal{P}_{\rm s}(k) = \log A_{\mathrm{s}} + \left(n_{\mathrm{s}}-1\right) \log \left(\frac{k}{k_{\star}}\right) ,
	\label{PLS}
\end{equation}
where we fix $k_{\star}=0.05$ Mpc$^{-1}$. We keep the amplitude of the spectrum $A_{\mathrm{s}}$ a free parameter, while we compute the spectral index $n_s$ in terms of the number of e-folds $N$ by using Eq.~\eqref{eq.ns}. Similarly, for primordial gravitational waves, we assume a power-law spectrum
\begin{equation}
	\log \mathcal{P}_{\rm T}(k) = \log (A_{\mathrm{T}}) + n_{\rm T} \log \left(\frac{k}{k_{\star}}\right) , 
	\label{PLT}
\end{equation}
computing the amplitude $r\equiv \mathcal P_{\rm T} (k_{\star})/ \mathcal P_{\rm S}(k_{\star})$ in terms of the number of e-folds $N$ and the additional parameter $\alpha$ by means of Eq.~\eqref{eq.r}. As for the tensor tilt $n_{\rm T}$, we assume the usual slow-roll relation $n_{\rm T}=-r/8$.

In $\alpha$-attractor quintessential inflation, the parameters related to the DE equation of state $w_0$ and $w_a$ -- which describe the dynamical evolution of the second phase of accelerated expansion of the universe -- can be determined starting from the same inflationary parameters (or, alternatively, from $\alpha$) through Eq.~\eqref{eq.w0} and Eq.~\eqref{eq.wa}. Therefore, within the parameterization resulting from this model, the entire evolutionary history of the universe can be completely described by specifying only 7 free parameters: the energy density of baryonic matter ($\Omega_b\, h^2$), the density of cold dark matter ($\Omega_c\, h^2$), the optical density during reionization ($\tau_\mathrm{reio}$), the current rate of expansion of the Universe ($H_0$) -- or equivalently the angular size of the sound horizon ($\theta_{\rm{MC}}$),  the amplitude of the spectrum of primordial perturbations ($A_s$)  and two additional free parameters of the model, namely the number of e-folds $N$ and $\alpha$. 
Note there are the same number of parameters as in $\Lambda$CDM. 

We compute the theoretical model using the Boltzmann integrator code \texttt{CAMB}~\cite{Lewis:1999bs,Howlett:2012mh} while we explore the parameter space of our models by means of the publicly available sampler \texttt{COBAYA}~\cite{Torrado:2020dgo}. The code explores the posterior distributions of a given parameter space using the Monte Carlo Markov Chain (MCMC) sampler developed for \texttt{CosmoMC}~\cite{Lewis:2002ah} and tailored for parameter spaces with speed hierarchy implementing the ``fast dragging'' procedure developed in~\cite{Neal:2005}.

Our baseline data-sets consist of:

\begin{itemize}
	
\item The full Planck 2018 temperature and polarization (TT TE EE) likelihoods~\cite{Planck:2019nip,Planck:2018vyg,Planck:2018nkj} in combination with the Planck 2018 lensing likelihood~\cite{Planck:2018lbu}, reconstructed from the temperature 4-point correlation function.  We refer to this dataset as \textbf{P18}.

\item The Atacama Cosmology Telescope temperature and polarization (TT TE EE) ACTPol-DR4 likelihood~\citep{ACT:2020frw}, in combination with the recent ACTPol-DR6 lensing likelihood~\cite{ACT:2023kun}. We refer to this dataset as \textbf{ACT}.

\item The 2018 B-modes polarization likelihood from the BICEP/Keck Collaboration~\cite{BICEP:2021xfz}. We refer to this dataset as \textbf{BK18}. 

\item Baryon Acoustic Oscillation (BAO) and Redshift-Space Distortions (RSD) measurements from the completed SDSS-IV eBOSS survey. These include isotropic and anisotropic distance and expansion rate measurements, as well as measurements of $f\sigma_8$, and are summarized in Table~3 of Ref.~\cite{eBOSS:2020yzd}. We refer to this dataset as \textbf{BAO} (though it also includes RSD). 

\item 1701 light curves for 1550 distinct SNeIa in the redshift range $0.001 < z < 2.26$ collected in the \textit{PantheonPlus} sample~\cite{Scolnic:2021amr}. 
We refer to this dataset as \textbf{SN}.
	
\end{itemize}
The convergence of the chains obtained with this procedure is tested using the Gelman-Rubin criterion~\cite{Gelman:1992zz} using a threshold for chain convergence of $R-1 \lesssim 0.01$.


\section{Results for Planck}  \label{sec:planck} 

In this section, we present the results obtained by focusing on CMB measurements released by the Planck satellite, combined with other datasets listed in \autoref{sec:method}. In particular, we consider three different scenarios of quintessential inflation, assuming different priors on the parameter $\alpha$. Therefore, we divide this section as follows: In \autoref{sec:P18_free}, we discuss the most general case where this parameter is free to vary over a wide uninformative range $\alpha \in [0,\alpha_{\rm max}]$ with $\alpha_{\rm max}\gg 1$.  Instead, in \autoref{sec:P18_73}, we focus on a continuous range $\alpha=[0,7/3]$, inspired by models of supergravity where $\alpha$, being related to Poincaré disks of the symmetry group, can take on (discrete) values up to $\alpha_{\rm max}=7/3$. Finally, in \autoref{sec:P18_1}, we fix $\alpha=1$, corresponding to Starobinsky and Higgs inflation.

\subsection{$\alpha\in [0\,,\,\alpha_{\rm max}\gg 1]$}
\label{sec:P18_free}

\begin{figure*}
    \centering
    \includegraphics[width=1\linewidth]{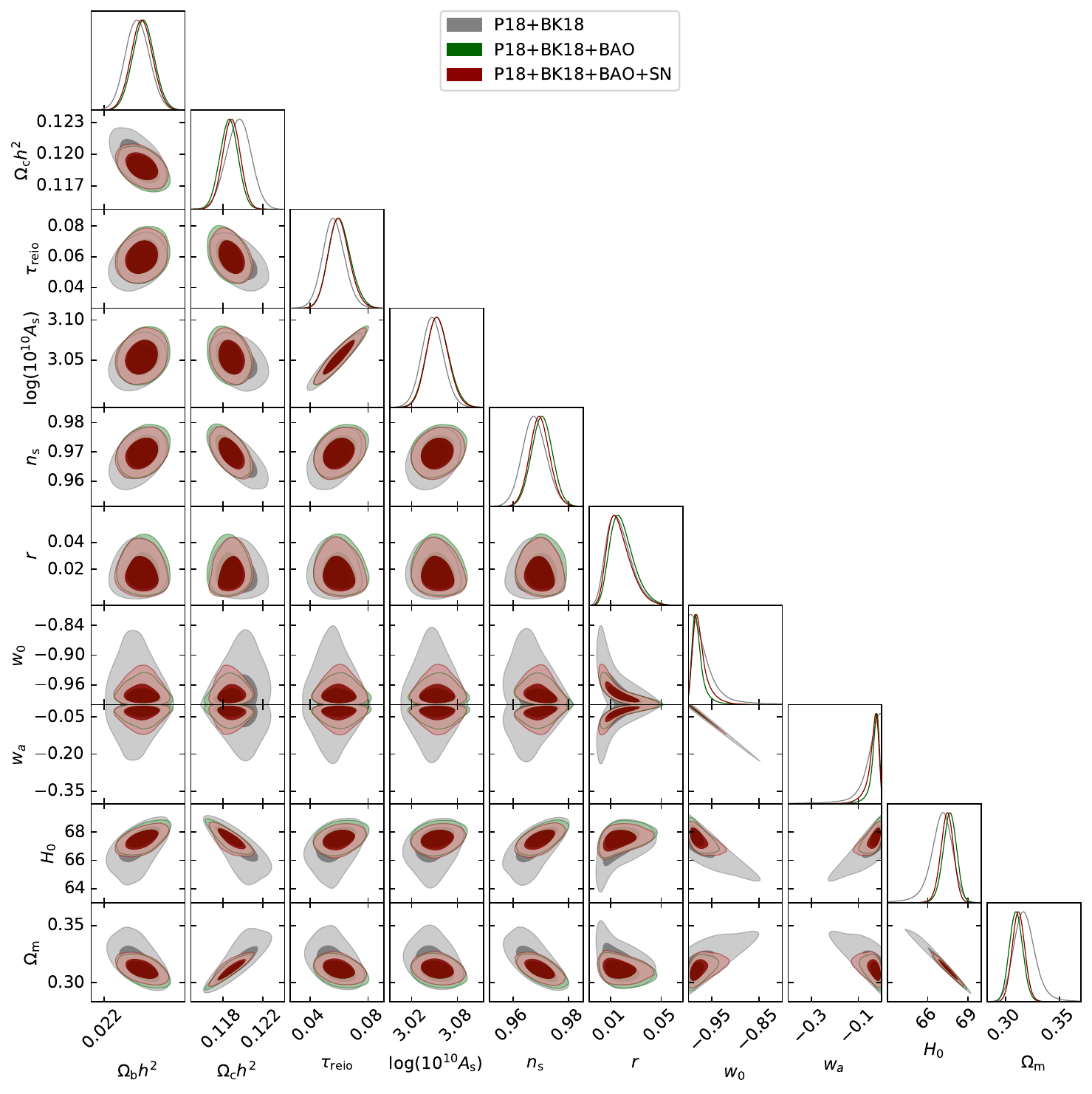}
    \caption{One-dimensional posterior probability distribution functions and two-dimensional contours at 68\% and 95\% CL for the parameters of interest in $\alpha$-attractor quintessential inflation. We consider three reference datasets indicated in the legend and adopt a prior $\alpha \in [0, \alpha_{\rm max} \gg 1]$.}
    \label{fig:Planck_alpha_free}
\end{figure*}

\begin{table*}[htpb!]
\begin{center}
\renewcommand{\arraystretch}{1.5}
\resizebox{0.85 \textwidth}{!}{
\begin{tabular}{l c c c l c c c c c c c c c c c }
\hline
\textbf{Parameter} & \textbf{ P18+BK18 } & \textbf{ P18+BK18+BAO } & \textbf{ P18+BK18+BAO+SN } & \textbf{Prior/Value}\\ 
\hline\hline
$ \boldsymbol{\Omega_\mathrm{b} h^2}  $  & $0.02241\pm 0.00015$ & $0.02248\pm 0.00013$ & $0.02246\pm 0.00013$ &$  [0.005\,,\,0.1]$\\ 
Bestfit:&[$0.022380$]&[$0.022534$]&[$0.022446$]&\\
\hline
$ \boldsymbol{\Omega_\mathrm{c} h^2}  $ & $0.1196\pm 0.0012$ & $0.11860\pm 0.00089$ & $0.11885\pm 0.00087$ &$  [0.001\,,\,0.1]$\\ 
Bestfit:&[$0.120385$]&[$0.119258$]&[$0.119496$]&\\
\hline
$ \boldsymbol{100\theta_\mathrm{MC}}  $ & $1.04097\pm 0.00031$ & $1.04108\pm 0.00029$ & $1.04106\pm 0.00029$ &$  [0.5\,,\,10]$\\
Bestfit:&[$1.040969$]&[$1.040933$]&[$1.041011$]&\\
\hline
$ \boldsymbol{\tau_\mathrm{reio}}  $    & $0.0563\pm 0.0076$ & $0.0604^{+0.0069}_{-0.0081}$ & $0.0600^{+0.0067}_{-0.0076}$ &$  [0.01\,,\,0.8]$\\ 
Bestfit:&[$0.052943$]&[$0.060925$]&[$0.057164$]&\\
\hline
$ \boldsymbol{\log(10^{10} A_\mathrm{s})}  $ & $3.048\pm 0.015$ & $3.054\pm 0.015$ & $3.054\pm 0.014$ &$  [1.61\,,\,3.91]$\\ 
Bestfit:&[$3.053140$]&[$3.064418$]&[$3.056218$]&\\
\hline
$\boldsymbol{N}$ & $63.0^{+6.1}_{-10}$ & $68.6^{+6.7}_{-10}$ & $66.6^{+5.9}_{-9.2}$ &$  [10\,,\,200]$\\ 
Bestfit:&[$56.916153$]&[$68.145933$]&[$61.063004$]&\\
\hline
$ \boldsymbol{\alpha}  $ & $5.8^{+1.6}_{-5.1}$ & $7.7^{+2.4}_{-5.3}$ & $6.3^{+1.9}_{-4.6}$ &$  [0\,,\,100]$\\
Bestfit:&[$1.523394$]&[$3.748671$]&[$3.217857$]&\\
\hline
$ \boldsymbol{n_\mathrm{s}}  $ & $0.9676\pm 0.0043$ & $0.9703\pm 0.0037$ & $0.9695\pm 0.0036$ \\
Bestfit:&[$0.964861$]&[$0.970651$]&[$0.967247$]&\\
\hline
$ \boldsymbol{r}  $ & $0.0168^{+0.0071}_{-0.011}$ & $0.0193^{+0.0069}_{-0.011}$ & $0.0171^{+0.0063}_{-0.011}$ \\ 
Bestfit:&[$0.005643$]&[$0.009687$]&[$0.010356$]&\\
\hline
$ \boldsymbol{w_0}  $ & $-0.963^{+0.016}_{-0.038}$ & $-0.9793^{+0.0061}_{-0.014}$ & $-0.9745^{+0.0077}_{-0.018}$ \\
Bestfit:&[$-0.927063$]&[$-0.970360$]&[$-0.965470$]&\\
\hline
$ \boldsymbol{w_a}  $ & $-0.056^{+0.057}_{-0.024}$ & $-0.0311^{+0.022}_{-0.0091}$ & $-0.038^{+0.028}_{-0.012}$ \\ 
Bestfit:&[$-0.109405$]&[$-0.044460$]&[$-0.051794$]&\\
\hline
$ \boldsymbol{\Omega_m} $ & $0.3188^{+0.0081}_{-0.012}$ & $0.3099\pm 0.0062$ & $0.3122\pm 0.0059$ \\ 
Bestfit:&[$0.329243$]&[$0.315068$]&[$0.317412$]&\\
\hline
$ \boldsymbol{H_0 } $ & $66.92^{+0.98}_{-0.62}$ & $67.63^{+0.52}_{-0.47}$ & $67.44\pm 0.49$ \\ 
Bestfit:&[$65.998056$]&[$67.237179$]&[$67.023797$]&\\
\hline
$ \boldsymbol{\sigma_8}$  & $0.8058^{+0.0084}_{-0.0058}$ & $0.8084\pm 0.0064$ & $0.8082\pm 0.0065$ \\ 
Bestfit:&[$0.804539$]&[$0.813136$]&[$0.809256$]&\\
\hline 
$\Delta \chi^2_{\texttt{low-TT}}$ & $-0.18$ & $+0.12$ & $+0.09$ \\ 
$\Delta \chi^2_{\texttt{low-EE}}$ & $+0.14$ & $+1.54$ & $+0.70$\\ 
$\Delta \chi^2_{\texttt{TTTEEE}}$ & $-1.96$ & $-0.42$ & $+0.43$ \\
$\Delta \chi^2_{\texttt{lensing}}$ & $+0.20$ & $+0.06$ & $-0.45$\\
$\Delta \chi^2_{\texttt{BK18}}$ & $-0.23$ & $-1.19$ & $+0.71$\\
$\Delta \chi^2_{\texttt{BAO}}$ & -- & $-0.81$ & $-0.30$\\ 
$\Delta \chi^2_{\texttt{SN}}$ & -- & --  & $-1.24$ \\ 
\hline 
$\Delta \chi^2_{\texttt{tot}}$ & $-2.03$ & $-0.70$ & $-0.06$ \\ 
\hline \hline
\end{tabular} }
\end{center}
\caption{Results for different combinations of data involving Planck-2018, BK18, BAO, and SN. The constraints on parameters are given at 68\% CL, while the upper/lower bounds are given at 95\% CL. For each parameter, the best-fit value is provided as well in brackets. Parameters are varied in the prior range (or fixed to the values) reported in the last column of the table. We provide the $\Delta \chi^2$ between the best fit for $\alpha$-attractor quintessential inflation and $\Lambda$CDM for each likelihood used in the analysis. }
\label{tab.results.P18.alpha_free}
\end{table*}

The results obtained by allowing $\alpha$ to vary over a wide range are summarized in \autoref{tab.results.P18.alpha_free}, along with the flat priors adopted for all free parameters. The marginalized one-dimensional probability distribution functions and the two-dimensional correlations among different parameters are shown in \autoref{fig:Planck_alpha_free} for various combinations of data.

By focusing solely on temperature and polarization CMB measurements released by the Planck satellite, in combination with B-mode polarization measurements from the Bicep/Keck collaboration, we are able to derive precise constraints on all parameters. When comparing our results to those documented in the literature, where all quantities are treated independently, it becomes apparent that assuming the theoretical relations predicted by the quintessential $\alpha$-attractor allows us to establish more stringent constraints on both inflationary parameters and those linked to the DE equation of state. For the number of e-folds $N$ and the parameter $\alpha$, we obtain $N = 63^{+6.1}_{-10}$ and $\alpha = 5.8^{+1.6}_{-5.1}$ both at the 68\% confidence level (hereafter, we will quote all the two-tail constraints on parameters at the 68\% confidence level and all upper/lower bounds at the 95\% confidence level, without repeating this information each time). These constraints translate into a bound on the spectral index $n_s=0.9676 \pm 0.0043$, while for the DE parameters, we obtain $w_0=-0.963^{+0.016}_{-0.038}$ and $w_a=-0.056^{+0.057}_{-0.024}$. A point that certainly deserves to be highlighted is that the interdependence of these parameters gives rise to testable predictions. An illustrative example supporting this claim is the authentic prediction of a non-zero value for the tensor amplitude $r=0.0168^{+0.0071}_{-0.011}$. Notice that this result (as well as all results on $r$ discussed in this work) should not be interpreted as direct observational evidence of a non-zero tensor amplitude in the data. In fact, as it is well known, analyzing the standard $\Lambda$CDM+$r$ case (thus without assuming any model of inflation or quintessential inflation), the detection of primordial tensor modes remains elusive, and the combined analysis of Planck and BK18 data only sets an upper bound of $r < 0.037$ at a 95\% CL~\cite{BICEP:2021xfz}. That being said, in $\alpha$-attractor quintessential inflation, $r$ is linked to $n_s$ and the late-time parameters $w_0$ and $w_a$. Therefore, constraints on these parameters translate into constraints on $r$. Within this model, given the joint constraints on the different early and late-time parameters, $r$ should lie in this interval at a certain confidence level. Clearly, the results obtained on both inflationary parameters and those associated with the late-time evolution of the universe represent significant joint predictions that can undergo testing through future CMB and large-scale structure experiments, ultimately providing support for or ruling out the model.

As a next step, we incorporate into the analysis large scale structure measurements 
in the form of BAO, always considering them in conjunction with Planck and BK18 data. The inclusion of BAO significantly enhances our constraining power, as we become more sensitive to the effects resulting from the dynamical behavior of DE at late times. As a matter of fact, much tighter constraints are obtained on $w_0=-0.9793^{+0.0061}_{-0.014}$ and $w_a=-0.0311^{+0.022}_{-0.0091}$. From  \autoref{fig:Planck_alpha_free} it is easy to see that the contours shrink around values closer to the standard cosmological model (i.e., $w_0 = -1$ and $w_a = 0$), reinforcing the general preference of BAO for an equation of state consistent with a cosmological constant. Given the link between inflation and DE, the increased ability to constrain the late-time Universe also influences the constraints on the inflationary Universe. Specifically, the refined constraints on the DE equation of state translate into a preference for a higher value of the number of e-folds, $N=68.6^{+6.7}_{-10}$, leading to a larger scalar tilt $n_s=0.9703\pm 0.0037$. In contrast, predictions for the tensor amplitude $r=0.0193^{+0.0069}_{-0.011}$ remain nearly unchanged, as higher values of $N$ are counterbalanced by larger values of $\alpha=7.7^{+2.4}_{-5.3}$.

Finally, we consider the combination P18+BK18+BAO+SN. Incorporating luminosity distance measurements obtained through supernova observations yields results similar to those discussed so far, although the error bars on most parameters appear to be slightly reduced. However, from \autoref{fig:Planck_alpha_free}, it is evident that all model predictions for key quantities such as $n_s=0.9695\pm 0.0036$, $r=0.0171^{+0.0063}_{-0.011}$, $w_0=-0.9745^{+0.0077}_{-0.018}$, and $w_a=-0.038^{+0.028}_{-0.012}$ remain essentially unchanged.

As a final important test, we assess whether the model can precisely fit the various observations considered in the analysis. To do so, for each given dataset $\mathcal D$, in \autoref{tab.results.P18.alpha_free} we present the corresponding $\Delta\chi^2_{\mathcal D}$ computed with respect to the standard cosmological model:
\begin{equation}
\Delta\chi^2_{\mathcal D} = \chi^2_{\mathcal D} - \chi^2_{\mathcal D,\Lambda\text{CDM}}\ ,
\label{eq:chi2}
\end{equation}
where $\chi^2_{\mathcal D}$ is the minimum $\chi^2$ obtained for the dataset $\mathcal D$ within quintessential $\alpha$-attractor inflation, and $\chi^2_{\mathcal D,\Lambda\text{CDM}}$ is the minimum $\chi^2$ obtained for the same dataset within $\Lambda$CDM (the results for $\Lambda$CDM, as well as $\chi^2_{\mathcal D, \Lambda\text{CDM}}$, are presented in the \hyperref[sec:supmat]{supplementary material} for the same combinations of data). Notice that $\Delta\chi^2$ is defined in such a way that negative (positive) values represent improvements (worsening) in the fit compared to $\Lambda$CDM. Across all three data combinations in the table, we consistently observe a marginal improvement in $\Delta\chi^2$, demonstrating that the model can provide a fit to data as good as that achieved within the standard cosmological model.

\subsection{$\alpha\in [0\,,\,7/3]$}
\label{sec:P18_73}

\begin{figure*}
    \centering
    \includegraphics[width=1\linewidth]{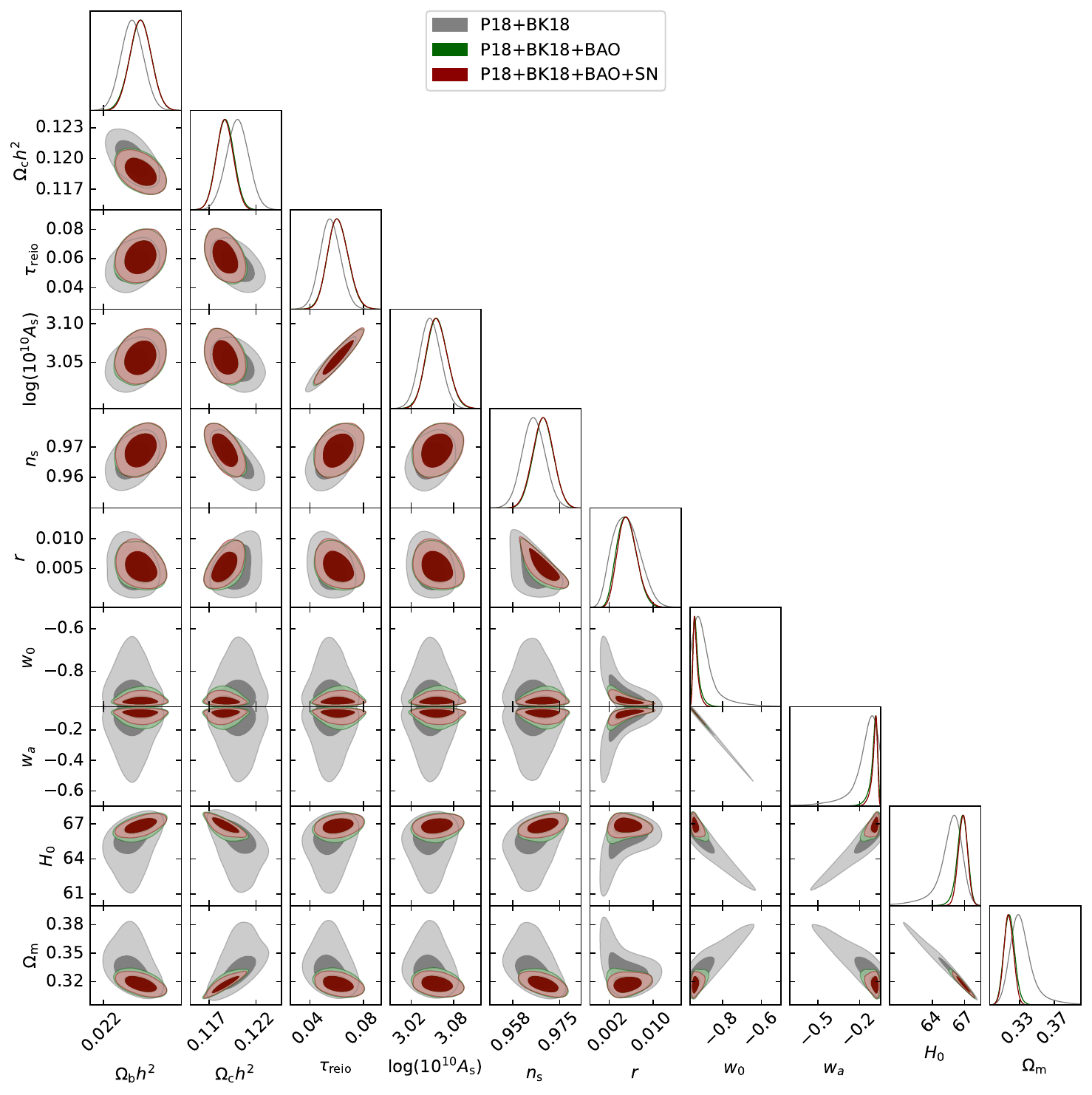}
    \caption{One-dimensional posterior probability distribution functions and two-dimensional contours at 68\% and 95\% CL for the parameters of interest in $\alpha$-attractor quintessential inflation. We consider three reference datasets indicated in the legend and adopt a supergravity-inspired prior $\alpha \in [0,7/3]$.}
    \label{fig:Planck_alpha_73}
\end{figure*}

\begin{table*}[htpb!]
\begin{center}
\renewcommand{\arraystretch}{1.5}
\resizebox{0.85 \textwidth}{!}{
\begin{tabular}{l c c c l c c c c c c c c c c c }
\hline
\textbf{Parameter} & \textbf{ P18+BK18 } & \textbf{ P18+BK18+BAO } & \textbf{ P18+BK18+BAO+SN } & \textbf{Prior/Value}\\ 
\hline\hline

$ \boldsymbol{\Omega_\mathrm{b} h^2}  $ & $0.02237\pm 0.00014$ & $0.02248\pm 0.00014$ & $0.02248\pm 0.00013$ &$  [0.005\,,\,0.1]$\\ 
Bestfit:&[$0.022478$]&[$0.022555$]&[$0.022412$]&\\
\hline
$ \boldsymbol{\Omega_\mathrm{c} h^2}  $ & $0.1200\pm 0.0012$ & $0.11872\pm 0.00093$ & $0.11868\pm 0.00089$ &$  [0.001\,,\,0.1]$\\ 
Bestfit:&[$0.119982$]&[$0.118382$]&[$0.118677$]&\\
\hline
$ \boldsymbol{100\theta_\mathrm{MC}}  $ & $1.04091\pm 0.00030$ & $1.04108\pm 0.00030$ & $1.04107\pm 0.00029$ &$  [0.5\,,\,10]$\\ 
Bestfit:&[$1.040871$]&[$1.040984$]&[$1.041175$]&\\
\hline
$ \boldsymbol{\tau_\mathrm{reio}}  $ & $0.0553\pm 0.0076$ & $0.0610\pm 0.0077$ & $0.0611^{+0.0071}_{-0.0081}$ &$  [0.01\,,\,0.8]$\\ 
Bestfit:&[$0.057438$]&[$0.061664$]&[$0.064106$]&\\
\hline
$ \boldsymbol{\log(10^{10} A_\mathrm{s})}  $ & $3.047\pm 0.015$ & $3.056\pm 0.015$ & $3.057^{+0.014}_{-0.016}$ &$  [1.61\,,\,3.91]$\\ 
Bestfit:&[$3.056608$]&[$3.057068$]&[$3.063715$]&\\
\hline
$\boldsymbol{N}$ & $58.9^{+5.5}_{-8.3}$ & $65.5^{+6.1}_{-9.1}$ & $65.4^{+6.2}_{-9.2}$ &$  [10\,,\,200]$\\ 
Bestfit:&[$61.178771$]&[$68.034857$]&[$60.336977$]&\\
\hline
$ \boldsymbol{\alpha}  $ & $> 0.424$ & $> 1.04$ & $> 1.18$ &$  [0\,,\,7/3]$\\ 
Bestfit:&[$2.038146$]&[$2.198705$]&[$2.273368$]&\\
\hline
$ \boldsymbol{n_\mathrm{s}}  $ & $0.9655\pm 0.0041$ & $0.9690\pm 0.0037$ & $0.9690\pm 0.0037$ \\
Bestfit:&[$0.967309$]&[$0.970603$]&[$0.966853$]&\\
\hline
$ \boldsymbol{r}  $ & $0.0051^{+0.0021}_{-0.0028}$ & $0.0052^{+0.0016}_{-0.0019}$ & $0.0054^{+0.0015}_{-0.0019}$ \\
Bestfit:&[$0.006535$]&[$0.005700$]&[$0.007493$]&\\
\hline
$ \boldsymbol{w_0}  $ & $-0.895^{+0.027}_{-0.074}$  & $-0.9334^{+0.0084}_{-0.021}$ & $-0.9366^{+0.0072}_{-0.017}$ \\ 
Bestfit:&[$-0.945484$]&[$-0.949465$]&[$-0.951125$]&\\
\hline
$ \boldsymbol{w_a}  $ & $-0.158^{+0.11}_{-0.041}$ & $-0.0999^{+0.031}_{-0.013}$  & $-0.095^{+0.025}_{-0.011}$ \\ 
Bestfit:&[$-0.081774$]&[$-0.075802$]&[$-0.073313$]&\\
\hline
$ \boldsymbol{\Omega_m} $ & $0.3332^{+0.0079}_{-0.016}$ & $0.3182^{+0.0058}_{-0.0066}$ & $0.3174\pm 0.0055$ \\ 
Bestfit:&[$0.323674$]&[$0.313517$]&[$0.315015$]&\\
\hline
$ \boldsymbol{H_0 } $  & $65.6^{+1.4}_{-0.61}$ & $66.77^{+0.54}_{-0.44}$ & $66.84\pm 0.43$ \\ 
Bestfit:&[$66.492485$]&[$67.200828$]&[$67.076766$]&\\
\hline
$ \boldsymbol{\sigma_8}  $ & $0.796^{+0.013}_{-0.0063}$ & $0.8021\pm 0.0066$ & $0.8024\pm 0.0064$ \\ 
Bestfit:&[$0.807680$]&[$0.803760$]&[$0.807374$]&\\
\hline 
$\Delta \chi^2_{\texttt{low-TT}}$ & $-0.55$ & $-0.04$ & $+0.30$ \\ 
$\Delta \chi^2_{\texttt{low-EE}}$ & $+1.01$ & $+1.69$ & $+3.20$ \\ 
$\Delta \chi^2_{\texttt{TTTEEE}}$ & $-1.74$ & $+1.22$ & $+0.12$ \\
$\Delta \chi^2_{\texttt{lensing}}$ & $+0.11$ & $-0.11$ & $-0.64$ \\
$\Delta \chi^2_{\texttt{BK18}}$ & $-0.25$ & $-1.31$ & $-0.28$\\
$\Delta \chi^2_{\texttt{BAO}}$ & -- & $-1.21$ & $-1.53$ \\ 
$\Delta \chi^2_{\texttt{SN}}$ & -- & -- & $-1.37$ \\ 
\hline 
$\Delta \chi^2_{\texttt{tot}}$ & $-1.42$ & $+0.24$ & $-0.20$ \\ 
\hline \hline
\end{tabular} }
\end{center}
\caption{Results for different combinations of data involving  Planck-2018, BK18, BAO, and SN. The constraints on parameters are given at 68\% CL, while the upper/lower bounds are given at 95\% CL. For each parameter, the best-fit value is provided as well. Parameters are varied in the prior range (or fixed to the values) reported in the last column of the table. We provide the $\Delta \chi^2$ between the best fit for $\alpha$-attractor quintessential inflation and $\Lambda$CDM for each likelihood used in the analysis. }
\label{tab.results.P18.alpha_73}
\end{table*}

As a second step, we examine the same combinations of data, narrowing the prior for $\alpha$ to the range $\alpha\in[0,7/3]$. The results obtained in this case are reported in \autoref{tab.results.P18.alpha_73}, while the one-dimensional probability distributions and correlations among different parameters are shown in \autoref{fig:Planck_alpha_73}.

Reducing the prior range for $\alpha$ has a significant impact on the conclusions we can draw from the analysis of P18 and BK18 data. Focusing on inflationary parameters, we obtain a slightly lower number of e-folds $N=58.9^{+5.5}_{-8.3}$, which nevertheless remains largely in agreement with the values expected for this parameter to successfully describe the inflationary Universe, as well as with the results obtained considering a wider prior on $\alpha$. Regarding this latter parameter, we now obtain a lower limit $\alpha>0.424$ which clearly reflects the choice of adopting a more restrictive prior (note the best fit is $\alpha\approx2$). The same argument applies to the difference observed in the prediction for the amplitude of primordial gravitational waves. Referring back to Eq.~\eqref{eq.r}, it is easy to see that if $N$ does not change significantly, limiting the value of $\alpha$ implies confining ourselves to models that predict a lower $r$. For this reason, when performing the full MCMC analysis, we get $r=0.0051^{+0.0021}_{-0.0028}$ which is much lower than what was discussed in the previous subsection. This unequivocally demonstrates that models in which $\alpha$ is correlated with the Poincaré disks of the symmetry group and can only assume discrete values up to $\alpha_{\rm max}=7/3$ lead to predictions for the tensor amplitude that are challenging to measure directly. This range of $r$ (note best fits are $r\approx0.006$) is a major goal of future CMB experiments. As for the parameters governing the evolution of the Universe in later epochs, we note that the differences introduced by limiting the prior of $\alpha$ recast into $w_0=-0.895^{+0.027}_{-0.074}$ and $w_a=-0.158^{+0.11}_{-0.041}$, namely a strong preference for dynamical DE in the quintessence regime (note that $|w_a|$ is approximately double the value obtained by leaving $\alpha$ free to vary).

A similar dynamical DE component impacts the evolutionary history of the Universe at later times, leaving imprints in its related observables that we can test with current data. Therefore, as before, we introduce BAO measurements in the analysis to cover later cosmic epochs and gain some constraining power on cosmological parameters. As extensively documented in the literature, BAO data align well with a cosmological constant. Therefore, when this dataset is considered, $w_0 = -0.9334^{+0.0084}_{-0.021}$ and $w_a = -0.0999^{+0.031}_{-0.013}$ shift significantly towards the direction of the standard cosmological model. Given the interconnection of the two cosmological epochs, this preference for a $\Lambda$CDM-like late-time cosmology produces changes in the constraints on the inflationary parameters as well. For instance, we observe a shift in the number of e-folds $N = 65.5^{+6.1}_{-9.1}$, while the limit on $\alpha > 1.04$ moves above the Starobinsky inflation value of $\alpha=1$. However, these differences in $N$ and $\alpha$ compensate for each other, with negligible effects on the predictions for the tensor amplitude ($r = 0.0052^{+0.0016}_{-0.0019}$).

Including supernova measurements, our conclusions remain unchanged. By considering P18, BK18, BAO, and SN data altogether, the main difference we observe is a tiny reduction in the uncertainties for the vast majority of parameters, particularly $w_0 = -0.9366^{+0.0072}_{-0.017}$ and $w_a = -0.095^{+0.025}_{-0.011}$. Concerning inflationary parameters, despite a further shift in $\alpha > 1.18$, including SN measurements leaves the results discussed so far largely unaffected.

We assess whether restricting $\alpha\in[0,7/3]$ the model still provides an adequate description of all the datasets involved in the analysis. Specifically, following the previous subsection, we compare it to the standard cosmological model and evaluate the $\chi^2$ in both cases for the same combinations of data. Our results reveal minimal improvements in the $\chi^2$ value for P18+BK18 and for P18+BK18+BAO+SN, while a minor worsening is observed for P18+BK18+BAO. Overall, these differences are not substantial, leading us to the overarching conclusion that the model's ability to explain current observations remains comparable to $\Lambda$CDM.

\subsection{$\alpha=1$}
\label{sec:P18_1}

\begin{figure*}
    \centering
    \includegraphics[width=1\linewidth]{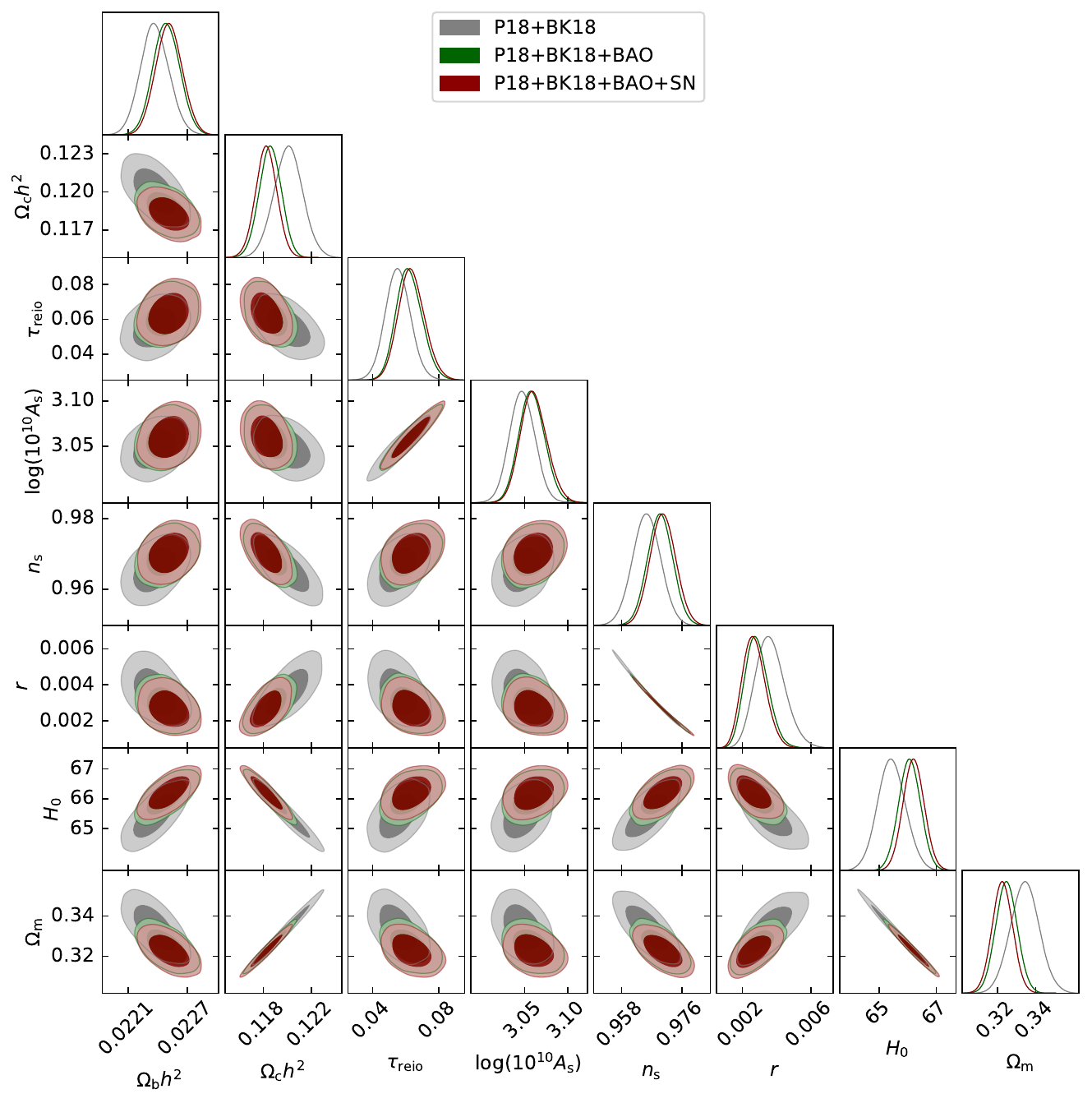}
    \caption{One-dimensional posterior probability distribution functions and two-dimensional contours at 68\% and 95\% CL for the parameters of interest in $\alpha$-attractor quintessential inflation. We consider three reference datasets indicated in the legend fixing $\alpha=1$ (and so $w_0\simeq -0.89$, $w_a=-0.16$ bt Eq.~\eqref{eq.w0} and Eq.~\eqref{eq.wa}).}
    \label{fig:Planck_alpha_1}
\end{figure*}

\begin{table*}[htpb!]
\begin{center}
\renewcommand{\arraystretch}{1.5}
\resizebox{0.85 \textwidth}{!}{
\begin{tabular}{l c c c l c c c c c c c c c c c }
\hline
\textbf{Parameter} & \textbf{ P18+BK18 } & \textbf{ P18+BK18+BAO } & \textbf{ P18+BK18+BAO+SN } & \textbf{Prior/Value}\\ 
\hline\hline

$ \boldsymbol{\Omega_\mathrm{b} h^2} $ & $0.02237\pm 0.00014$ & $0.02249\pm 0.00013$ & $0.02252\pm 0.00013$ &$  [0.005\,,\,0.1]$\\ 
Bestfit:&[$0.022504$]&[$0.022418$]&[$0.022494$]&\\
\hline
$ \boldsymbol{\Omega_\mathrm{c} h^2}  $ & $0.1201\pm 0.0012$ & $0.11857\pm 0.00091$ & $0.11824\pm 0.00088$ &$  [0.001\,,\,0.1]$\\ 
Bestfit:&[$0.119260$]&[$0.118064$]&[$0.118799$]&\\
\hline
$ \boldsymbol{100\theta_\mathrm{MC}}  $ & $1.04090\pm 0.00030$ & $1.04110\pm 0.00029$ & $1.04114\pm 0.00029$ &$  [0.5\,,\,10]$\\ 
Bestfit:&[$1.040867$]&[$1.041104$]&[$1.041049$]&\\
\hline
$ \boldsymbol{\tau_\mathrm{reio}}  $ & $0.0553\pm 0.0077$ & $0.0621^{+0.0070}_{-0.0082}$ & $0.0635^{+0.0073}_{-0.0083}$ &$  [0.01\,,\,0.8]$\\ 
Bestfit:&[$0.052012$]&[$0.059842$]&[$0.060756$]&\\
\hline
$ \boldsymbol{\log(10^{10} A_\mathrm{s})}  $ & $3.047\pm 0.015$ & $3.058\pm 0.015$ & $3.060^{+0.014}_{-0.016}$ &$  [1.61\,,\,3.91]$\\ 
Bestfit:&[$3.048496$]&[$3.060888$]&[$3.052388$]&\\
\hline
$\boldsymbol{N}$ & $58.8^{+5.5}_{-8.2}$ & $66.6^{+6.2}_{-9.5}$ & $68.6^{+6.6}_{-10}$ &$  [10\,,\,200]$\\ 
Bestfit:&[$55.505557$]&[$69.105377$]&[$62.982350$]&\\
\hline
$ \boldsymbol{\alpha} $ & -- & -- & -- & $\alpha=1$\\
\hline
$ \boldsymbol{n_\mathrm{s}}  $ & $0.9655\pm 0.0041$ & $0.9695\pm 0.0037$ & $0.9703\pm 0.0038$ \\
Bestfit:&[$0.963968$]&[$0.971059$]&[$0.968245$]&\\
\hline
$ \boldsymbol{r}  $ & $0.00363^{+0.00076}_{-0.00093}$ & $0.00283^{+0.00060}_{-0.00074}$ & $0.00268^{+0.00060}_{-0.00073}$ \\ 
Bestfit:&[$0.003895$]&[$0.002513$]&[$0.003025$]&\\
\hline
$ \boldsymbol{w_0}  $ & -- & -- & --  & $w_0\simeq-0.89$\\
\hline
$ \boldsymbol{w_a}  $ & -- & -- & --  & $w_a\simeq-0.16$ \\ 
\hline
$ \boldsymbol{\Omega_m} $ & $0.3343\pm 0.0075$ & $0.3247\pm 0.0055$ & $ 0.3226\pm 0.0053$ \\ 
Bestfit:&[$0.329125$]&[$0.322297$]&[$0.326075$]&\\
\hline
$ \boldsymbol{H_0 } $ & $65.43\pm 0.49$ & $66.06\pm 0.38$ & $66.21\pm 0.37$ \\ 
Bestfit:&[$65.779160$]&[$66.172437$]&[$65.976614$]&\\
\hline
$ \boldsymbol{\sigma_8}  $ & $0.7953\pm 0.0058$ & $0.7954\pm 0.0059$ & $0.7953^{+0.0057}_{-0.0064}$ \\ 
Bestfit:&[$0.792095$]&[$0.795467$]&[$0.793471$]&\\
\hline 
$\Delta \chi^2_{\texttt{low-TT}}$ & $+0.11$ & $-0.24$ & $-0.14$ \\ 
$\Delta \chi^2_{\texttt{low-EE}}$ & $-0.01$ & $+1.06$ & $+1.73$ \\ 
$\Delta \chi^2_{\texttt{TTTEEE}}$ & $-1.41$ & $+1.57$ & $+0.51$ \\
$\Delta \chi^2_{\texttt{lensing}}$ & $-0.22$ & $-0.04$ & $-0.45$ \\
$\Delta \chi^2_{\texttt{BK18}}$ & $-0.72$ & $+0.44$ & $-0.10$\\
$\Delta \chi^2_{\texttt{BAO}}$ & -- & $+0.57$ & $+1.84$ \\ 
$\Delta \chi^2_{\texttt{SN}}$ & -- & -- & $+0.45$ \\ 
\hline 
$\Delta \chi^2_{\texttt{tot}}$ & $-2.25$ & $+3.36$ & $+3.84$ \\ 
\hline \hline

\end{tabular} }
\end{center}
\caption{Results for different combinations of data involving  Planck-2018, BK18, BAO, and SN. The constraints on parameters are given at 68\% CL, while the upper/lower bounds are given at 95\% CL. For each parameter, the best-fit value is provided as well. Parameters are varied in the prior range (or fixed to the values) reported in the last column of the table. We provide the $\Delta \chi^2$ between the best fit for $\alpha$-attractor quintessential inflation and $\Lambda$CDM for each likelihood used in the analysis. }
\label{tab.results.P18.alpha_1}
\end{table*}

As a final case, we consider a model of quintessential inflation where $\alpha=1$ is fixed according to Starobinsky and Higgs inflation. In this scenario, the inflationary sector of the theory is described by a single degree of freedom, parameterized by the number of e-folds $N$. Additionally, when considering Eq.~\eqref{eq.w0} and Eq.~\eqref{eq.wa}, it is noteworthy that fixing $\alpha=1$ completely determines the parameters describing the dynamical behavior of the DE equation of state. This sets $w_0\simeq-0.89$ and $w_a\simeq-0.16$, deviating significantly from the cosmological constant. Consequently, although the total number of free parameters in this case is six (just like in $\Lambda$CDM with fixed $r=0$), the theory predicts a distinct late-time evolution of the Universe characterized by a dynamical quintessence DE component. Therefore, analyzing such a model can be intriguing to understand whether and to what extent it is effectively supported by current observations, which, in turn, impose significant constraints on late-time (new) physics.

The results are presented in \autoref{tab.results.P18.alpha_1} and depicted in \autoref{fig:Planck_alpha_1}. When considering the minimal combination of P18 and BK18 data, the number of e-folds $N=58.8^{+5.5}_{-8.2}$ remains consistent with values obtained earlier by varying $\alpha$. Consequently, no significant differences are observed for the spectral index. In contrast, in this scenario, we predict a tensor amplitude that, at the same $N$, is smaller by a factor of $\alpha$ compared to previous results: $r=0.00363^{+0.00076}_{-0.00093}$. Additionally, despite the fewer parameters involved in the model, leading to very precise predictions for $r$ (i.e., fixing $N$ via $n_s$ completely determines $r$), it is crucial to note that these predictions are identical to those of the Starobinsky model. Therefore, a joint analysis of early and late-time cosmology would still be necessary to provide evidence or rule out the model itself, just like in all the cases considered so far. As for the other parameters, most of them are in good agreement with the values obtained within $\Lambda$CDM. However, it is important to discuss a significant (albeit  expected) aspect. For the Hubble parameter we obtain $H_0 = 65.43 \pm 0.49$ km/s/Mpc, significantly lower than what is inferred within the standard cosmological model. This result increases the well-known Hubble tension~\cite{Bernal:2016gxb,Verde:2019ivm,DiValentino:2020zio,DiValentino:2021izs,Abdalla:2022yfr} between the value of the present-day expansion rate of the Universe derived from CMB data (that, as highlighted by this analysis, depends significantly on the theoretical framework we assume) and the result for the same parameter directly measured by the SH0ES collaboration using Cepheids and Type Ia supernovae ($H_0 = 73 \pm 1$ km/s/Mpc~\cite{Riess:2021jrx}). Truth be told, such a trend toward lower values of $H_0$ is also observed when $\alpha$ is varied within different prior ranges. Given the well-known degeneracy between parameters describing the DE equation of state (mainly $w_0$) and the Hubble constant, as extensively documented in the literature, in order to increase $H_0$ by altering the late-time cosmology, one needs to consider a phantom equation of state $w_0 < -1$. Conversely, a quintessential component has the opposite effect, reducing the value of $H_0$. Therefore, this preference for smaller $H_0$ can be seen as an expected peculiarity of any quintessential model of DE, with quintessential $\alpha$-attractor inflation not being an exception. That being said, varying $\alpha$ provides more freedom and the possibility to somewhat move toward the cosmological constant regime (formally recovered in the limit $\alpha \to \infty$) obtaining values of $H_0$ close to $\Lambda$CDM. Instead, fixing $\alpha = 1$ (and to some extent limiting $\alpha$ to the range $[0, 7/3]$), confines us to a completely quintessential region. In addition, the fewer numbers of free parameters reduce the error bars. As a result, we observe a genuine increase in the Hubble tension.

In this case, adding local universe observations (either in the form of BAO or SN measurements) is certainly interesting for several reasons. Firstly, the values of $w_0$ and $w_a$ are fixed by the model to non-standard values and cannot be altered in any way. Therefore, maintaining a good fit to low-redshift data could represent a considerable challenge. In this regard, we note that the inclusion of BAO or SN produces interesting results. Firstly, focusing on \autoref{fig:Planck_alpha_1}, we observe that the geometric degeneracy between $\Omega_m$ and $H_0$ (i.e., the fact that various combinations of $\Omega_m$ and $H_0$ can lead to the same value of the distance to the last scattering surface, and thus to the same value of the acoustic scale $\theta_s$, if the sound horizon $r_s$ remains unchanged~\cite{Bond:1997wr,Zaldarriaga:1997ch,Efstathiou:1998xx}) is partially overcome in the presence of late-time measurements. The reason is that CMB data are very sensitive to $\Omega_m h^2$, while both SN and BAO allow precise measurements of $\Omega_m$. Therefore, considering only CMB data, higher values of $\Omega_m$ could always be compensated by lower values of $H_0$, while adding late-time data disrupts this trend. As a result, BAO and SN exclude large values of $\Omega_m$, thereby increasing $H_0$ (although lower values are still obtained compared to $\Lambda$CDM). Given the positive correlation between $H_0$ and $n_s$, fixing $\Omega_m$ with low-redshift data leads to a shift in $n_s$ towards slightly higher values, $n_s \sim 0.97$.

However, the real question to answer is whether, by fixing $w_0 \simeq -0.89$ and $w_a \simeq -0.16$, the model is still appropriate for explaining observations. Comparing the best-fit $\chi^2$ with the one obtained in $\Lambda$CDM, we note that for P18+BK18, there is still an improvement. However, for the reasons highlighted earlier (particularly the geometric degeneracy among parameters), CMB measurements typically leave large room to accommodate new physics at late times. The true challenge comes from including low-redshift measurements. Not surprisingly, both considering BAO and SN, we notice a global worsening $\Delta\chi^2\sim 3 - 4$ compared to the standard cosmological model. This stems from the limited freedom we are left with, as none of the parameters describing the DE equation of state can be altered in any way. Therefore, based on current data, we can conclude that the deterioration of the fit and the worsening of the Hubble constant tension suggest that fixing $\alpha=1$ may not be the best choice for integrating inflation and DE within the quintessential $\alpha$-attractor models.\\

Overall, considering the three cases studied thus far, the best option based on Planck satellite data, BK18 likelihood, and low-redshift measurements seems to be to consider models that allow greater freedom in the value of the parameter $\alpha$. Values $\alpha > 1$ allow approaching $\Lambda$CDM at late-time (though still showing some dynamics) while not spoiling Inflation. Interestingly, they also lead to higher values of the tensor amplitude $r$ that could be visible by future CMB probes.


\section{Results for ACT}
\label{sec:ACT}

In this section, we consider the same 3 scenarios of
quintessential inflation already analyzed in light of Planck data, thus varying $\alpha \in [0\, , \, \alpha_{\rm max}]$ for $\alpha_{\rm max} \gg 1$ and $\alpha_{\rm max} = 7/3$ or fixing $\alpha=1$. However, now we use the temperature, polarization, and lensing measurements released by the Atacama Cosmology Telescope. 

Before proceeding further, we wish to elaborate on the reasons prompting us to replicate the same analysis for a different CMB experiment. Indeed, several persuasive aspects have led us to believe that reproducing the analysis for the ACT data is worthwhile. Firstly, this experiment is independent of Planck, offering us a significant opportunity to test previously obtained results with independent measurements. Secondly, ACT explores angular scales different from those measured by the Planck satellite. For instance, ACT has sensitivity in the spectrum of temperature anisotropies covering the multipole range $\ell \in [650, 4200]$. Therefore, it probes much smaller scales compared to Planck which for the same spectrum covers the multipole range $\ell \in [2, 2500]$. Since the effects of new physics may manifest themselves differently at different angular scales, considering small-scale CMB data can be crucial when testing models beyond $\Lambda$CDM. Interestingly, a few intriguing hints for new physics have emerged from ACT, as highlighted in a multitude of studies~\cite{Hill:2021yec,Kreisch:2022zxp,Brax:2023rrf,Brax:2023tvn,Giare:2023qqn}. Even sticking to the standard $\Lambda$CDM cosmology, the constraints on the 6 free parameters of the model obtained from ACT show some important differences from those derived by the Planck satellite~\cite{ACT:2020gnv}. Particularly relevant for our discussion is that this disagreement mainly involves inflation as Planck suggests $n_s \ne 1$ with a significance level of about $8.5\sigma$, whereas ACT indicates a preference for a scale-invariant Harrison-Zel'dovich spectrum ($n_s=1$)~\cite{ACT:2020gnv,Giare:2022rvg}, introducing tension between the two experiments that can be quantified at approximately $2.5\sigma$ Gaussian equivalent level~\cite{Handley:2020hdp,DiValentino:2022rdg}. 

In the context of the $\alpha$-attractor quintessential inflation, this discrepancy raises several interesting questions that we seek to explore. For instance, assuming a standard post-inflationary cosmology, models commonly favored by large-scale CMB measurements (particularly $\alpha$-attractor or Starobinsky inflation) somehow fail to explain small-scale CMB data~\cite{Giare:2023wzl}. This creates a dichotomy in the conclusions drawn about which model best describes current data when considering experiments probing different angular scales. One may wonder whether similar results extend to $\alpha$-attractor quintessential inflation. In this scenario, inflation not only shapes the early Universe but also influences its late-time evolution, posing the question of whether the preference for $n_s \sim 1$ observed in ACT remains upheld or whether it is possible to maintain a good fit to the ACT data while adjusting the value of the spectral index closer to the measurement by the Planck collaboration. In a broader context, we aim to study the implications of a scalar spectrum approaching the scale-invariant regime.
\subsection{$\alpha\in [0\,,\,\alpha_{\rm max}\gg 1]$}

\begin{figure*}
    \centering
    \includegraphics[width=1\linewidth]{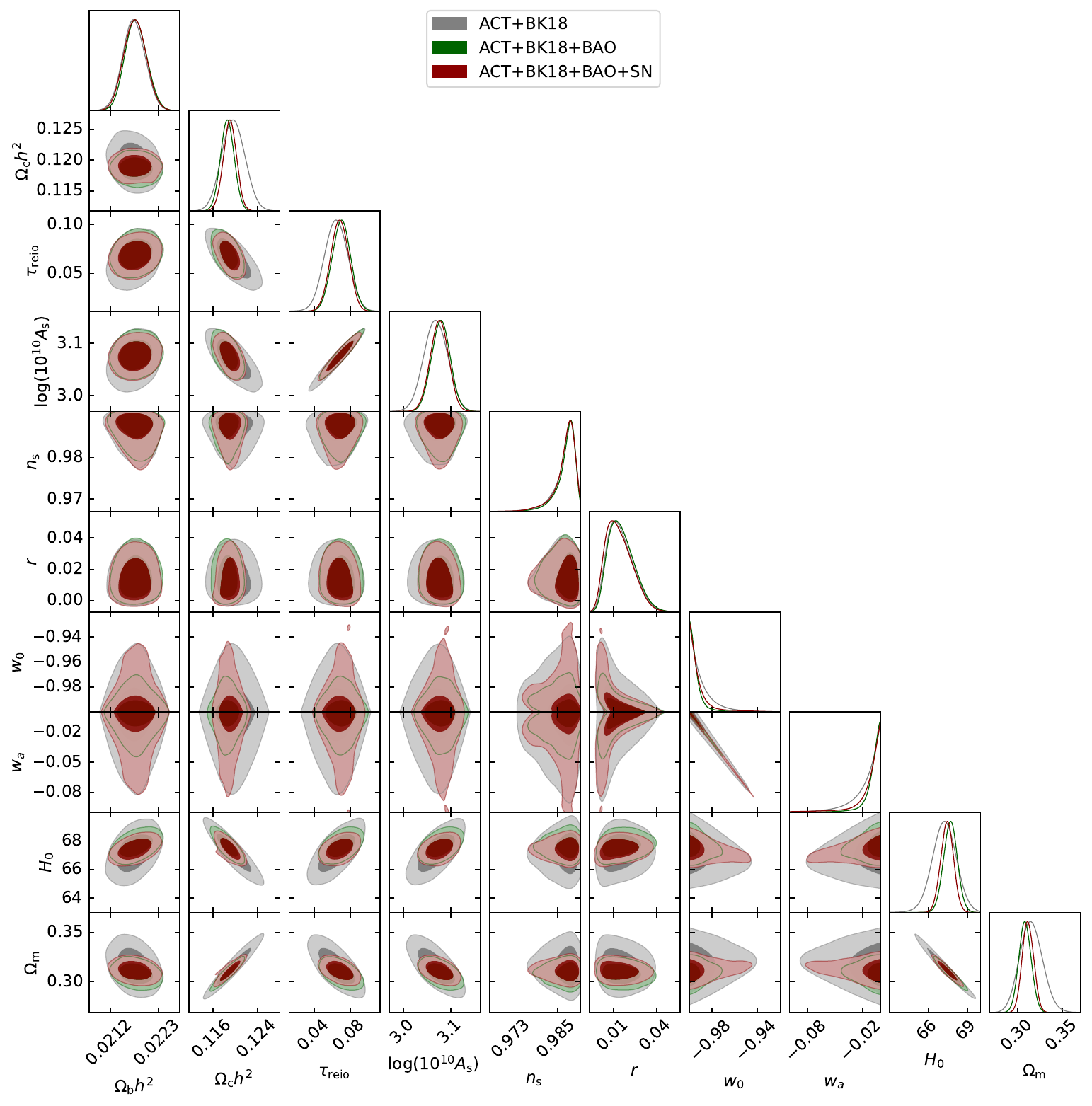}
    \caption{One-dimensional posterior probability distribution functions and two-dimensional contours at 68\% and 95\% CL for the parameters of interest in $\alpha$-attractor quintessential inflation. We consider three reference datasets indicated in the legend and adopt a prior $\alpha \in [0, \alpha_{\rm max} \gg 1]$.}
    \label{fig:ACT_alpha_free}
\end{figure*}

\begin{table*}[htpb!]
\begin{center}
\renewcommand{\arraystretch}{1.5}
\resizebox{0.85 \textwidth}{!}{
\begin{tabular}{l c c c l c c c c c c c c c c c }
\hline
\textbf{Parameter} & \textbf{ ACT+BK18 } & \textbf{ ACT+BK18+BAO } & \textbf{ ACT+BK18+BAO+SN } & \textbf{Prior/Value} \\ 
\hline\hline

$ \boldsymbol{\Omega_\mathrm{b} h^2}  $ & $0.02174\pm 0.00026$ &$ 0.02179\pm 0.00025$ & $0.02176\pm 0.00026$ &$  [0.005\,,\,0.1]$\\ 
Bestfit:&[$0.021748$]&[$0.021825$]&[$0.021877$]&\\
\hline
$ \boldsymbol{\Omega_\mathrm{c} h^2}  $ & $0.1197\pm 0.0021$ & $0.1186\pm 0.0012$ & $0.1190\pm 0.0012$ &$  [0.001\,,\,0.1]$\\ 
Bestfit:&[$0.119103$]&[$0.118734$]&[$0.117713$]&\\
\hline
$ \boldsymbol{100\theta_\mathrm{MC}}  $ & $1.04203\pm 0.00067$ & $1.04214\pm 0.00062$ & $1.04209\pm 0.00060$ &$  [0.5\,,\,10]$\\ 
Bestfit:&[$1.042179$]&[$1.042366$]&[$1.042224$]&\\
\hline
$ \boldsymbol{\tau_\mathrm{reio}}  $ & $0.064\pm 0.013$ & $0.070\pm 0.010$ & $0.0682\pm 0.0099$ &$ \mathcal{G}(0.065\,,\,0.015)$\\
Bestfit:&[$0.069651$]&[$0.071363$]&[$0.081290$]&\\
\hline
$ \boldsymbol{\log(10^{10} A_\mathrm{s})}  $ & $3.068\pm 0.025$ & $3.078\pm 0.019$ & $3.075\pm 0.019$ &$  [1.61\,,\,3.91]$\\ 
Bestfit:&[$3.074046$]&[$3.083627$]&[$3.098476$]&\\
\hline
$\boldsymbol{N}$ & $> 107$ & $> 111$ & $> 106$&$  [10\,,\,200]$\\ 
Bestfit:&[$195.813910$]&[$173.832850$]&[$191.776030$]&\\
\hline
$ \boldsymbol{\alpha}  $ & $< 77.4$ & $< 80.6$ & $< 75.6$ &$  [0\,,\,100]$ \\ 
Bestfit:&[$1.365259$]&[$21.082788$]&[$2.136796$]&\\
\hline
$ \boldsymbol{n_\mathrm{s}}  $ & $0.9871^{+0.0031}_{-0.0013}$ & $0.9874^{+0.0028}_{-0.0011}$ & $0.9871^{+0.0031}_{-0.0012}$ \\ 
Bestfit:&[$0.989786$]&[$0.988495$]&[$0.989571$]&\\
\hline
$ \boldsymbol{r}  $ & $0.0147^{+0.0068}_{-0.011}$ & $0.0154^{+0.0069}_{-0.011}$ & $0.0137^{+0.0069}_{-0.011}$ \\ 
Bestfit:&[$0.000427$]&[$0.008372$]&[$0.000697$]&\\
\hline
$ \boldsymbol{w_0}  $ & $-0.98969^{+0.0060}_{-0.0084}$ & $-0.9935^{+0.0028}_{-0.0071}$ & $ -0.9905^{+0.0034}_{-0.012}$ \\ 
Bestfit:&[$-0.918615$]&[$-0.994730$]&[$-0.948001$]&\\
\hline
$ \boldsymbol{w_a}  $ & $-0.0155^{+0.013}_{-0.00091}$ & $-0.0098^{+0.011}_{-0.0042}$ & $-0.0143^{+0.018}_{-0.0051}$\\ 
Bestfit:&[$-0.122077$]&[$-0.007905$]&[$-0.077998$]&\\
\hline
$ \boldsymbol{\Omega_m} $ & $0.315^{+0.011}_{-0.014}$ & $0.3079\pm 0.0069$ & $0.3112\pm 0.0064$ \\ 
Bestfit:&[$0.323245$]&[$0.307630$]&[$0.309959$]&\\
\hline
$ \boldsymbol{H_0 } $& $67.2\pm 1.1$ & $67.68\pm 0.53$ & $67.43\pm 0.48$ \\ 
Bestfit:&[$66.161587$]&[$67.750116$]&[$67.262908$]&\\
\hline
$ \boldsymbol{\sigma_8}  $& $0.8285^{+0.0082}_{-0.0070}$ & $0.8300\pm 0.0072$ & $0.8297\pm 0.0071$ \\ 
Bestfit:&[$0.818662$]&[$0.833249$]&[$0.828021$]&\\
\hline 
$\Delta \chi^2_{\texttt{ACT-DR4}}$ & $+0.24$ & $+1.41$ & $+1.22$ \\ 
$\Delta \chi^2_{\texttt{ACT-DR6}}$ & $-0.68$ & $-0.38$ & $-1.12$ \\ 
$\Delta \chi^2_{\texttt{BK18}}$ & $+0.38$ & $ -0.29$ & $+0.11$\\
$\Delta \chi^2_{\texttt{BAO}}$ & -- & $+0.03$ & $-0.60$ \\ 
$\Delta \chi^2_{\texttt{SN}}$ & -- & -- & $-1.72$ \\ 
\hline 
$\Delta \chi^2_{\texttt{tot}}$ & $-0.06$ & $+0.77$ & $-2.11$ \\ 
\hline \hline
\end{tabular} }
\end{center}
\caption{Results for different combinations of data involving  ACT(DR4+DR6), BK18, BAO, and SN. The constraints on parameters are given at 68\% CL, while the upper/lower bounds are given at 95\% CL. For each parameter, the best-fit value is provided as well. Parameters are varied in the prior range (or fixed to the values) reported in the last column of the table. We provide the $\Delta \chi^2$ between the best fit for $\alpha$-attractor quintessential inflation and $\Lambda$CDM for each likelihood used in the analysis. }
\label{tab.results.ACT.alpha_free}
\end{table*}

The constraints obtained by varying $\alpha$ in the range $\alpha\in[0, \alpha_{\rm max}\gg 1]$ are presented in \autoref{tab.results.ACT.alpha_free} and \autoref{fig:ACT_alpha_free}. Focusing on the temperature, polarization, and lensing measurements released by ACT in combination with the BK18 B-mode polarization measurements, the first noticeable difference is that now we obtain a lower limit for the number of e-folds $N>107$ and an upper limit $\alpha<77.4$ (both at 95\% CL). As largely expected from previous considerations, the loss of constraining power on $N$ primarily stems from the ACT preference for a scale-invariant Harrison-Zel'dovich spectrum. From Eq.~\eqref{eq.ns} it is easy to see that $n_s=1$ is formally recovered in the limit $N\to \infty$, explaining why the constraints on $N$ shift toward such large values. On the other hand, large values $\alpha\gg 1$ have the effect of increasing the tensor amplitude $r$ while simultaneously driving the parameters describing the DE equation of state towards the cosmological constant regime $w_0\to-1$ and $w_a\to0$. Although one might expect the effects on the tensor amplitude to rule out exceedingly large values of $\alpha$ (as it happens analyzing the Planck data), it should be noted that in Eq.~\eqref{eq.r} large values of $\alpha$ can always be balanced by large values of $N$. In turn, because of Eq.~\eqref{eq.ns}, large values of $N$ lead to a shift in $n_s=0.9871^{+0.0031}_{-0.0013}$ towards a scale-invariant spectrum, which is strongly favored by ACT data. As a result, we observe a clear degeneracy between these two parameters, allowing for $\alpha\gg1$ without compromising the result on the tensor amplitude ($r=0.0147^{+0.0068}_{-0.011}$) that remains consistent with B-modes polarization measurements. Concerning the late-time effects produced by $\alpha\gg1$, as we move in the direction of a cosmological constant, we still remain in good agreement with ACT data, getting $w_0=-0.98969^{+0.0060}_{-0.0084}$ and $w_a=-0.0155^{+0.013}_{-0.00091}$. Notice also that restricting ourselves to a region of the parameter space close to a standard late-time $\Lambda$CDM cosmology allows us to recover familiar values for the Hubble parameter, $H_0=67.2\pm1.1$ km/s/Mpc.

Including low-redshift data in the analysis leads to a reduction in uncertainties for most parameters due to the various effects discussed in the preceding sections. These effects include breaking geometric degeneracy between $H_0$ and $\Omega_m$ (see also \autoref{fig:ACT_alpha_free}), as well as enhancing the constraint power on parameters describing the background evolution of the late-time Universe. However, since the ACT preference for a large $n_s$ already forces us very close to a $\Lambda$CDM late-time cosmology, including BAO or SN measurements has a marginal impact on the conclusions we can draw regarding the relevant parameters of the model. In fact, the same considerations pointed out in the CMB analysis remain fundamentally unchanged.

For the three studied datasets, we compare the $\chi^2$ value corresponding to the best fit obtained in quintessential $\alpha$-attractor inflation with that obtained within the standard $\Lambda$CDM model. We observe that for ACT+BK18(+BAO), the difference between the two models is not particularly significant in terms of improvements or deteriorations in the fit. On the brighter side, for ACT+BK18+BAO+SN, we observe a slight improvement, with $\Delta\chi^2\sim -2$ compared to $\Lambda$CDM. A substantial portion of this improvement appears to come from the lensing likelihood of ACT-DR6 and the SN measurements.
\subsection{$\alpha\in [0\,,\,7/3]$}

\begin{figure*}
    \centering
    \includegraphics[width=1\linewidth]{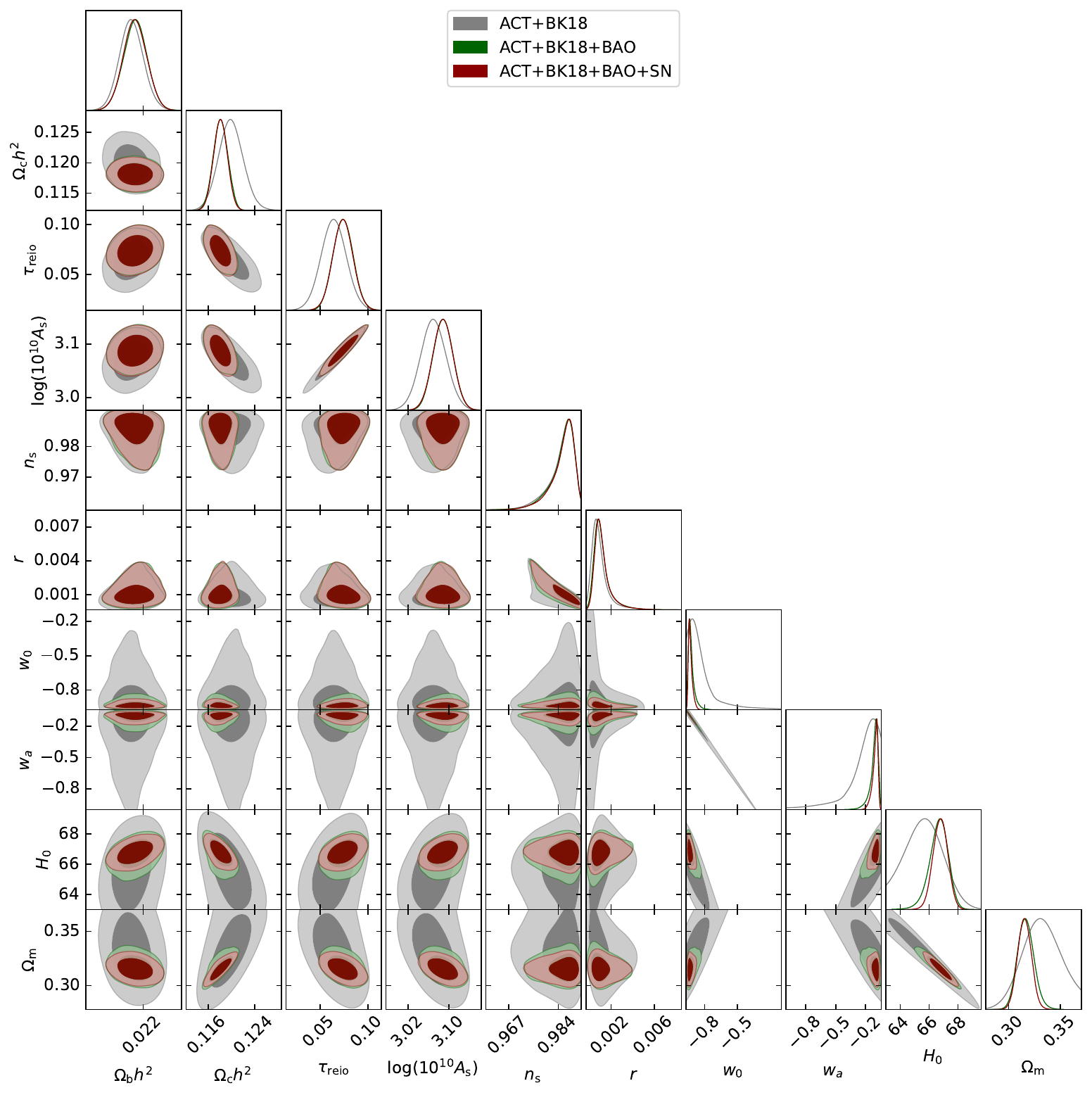}
    \caption{One-dimensional posterior probability distribution functions and two-dimensional contours at 68\% and 95\% CL for the parameters of interest in $\alpha$-attractor quintessential inflation. We consider three reference datasets indicated in the legend and adopt a supergravity-inspired prior $\alpha \in [0,7/3]$.}
    \label{fig:ACT_alpha_73}
\end{figure*}

\begin{table*}[htpb!]
\begin{center}
\renewcommand{\arraystretch}{1.5}
\resizebox{0.85 \textwidth}{!}{
\begin{tabular}{l c c c l c c c c c c c c c c c }
\hline
\textbf{Parameter} & \textbf{ ACT+BK18 } & \textbf{ ACT+BK18+BAO } & \textbf{ ACT+BK18+BAO+SN } & \textbf{Prior/Value} \\ 
\hline\hline

$ \boldsymbol{\Omega_\mathrm{b} h^2}  $ & $0.02174\pm 0.00026$ & $0.02183\pm 0.00026$ & $0.02182\pm 0.00026$ &$  [0.005\,,\,0.1]$ \\ 
Bestfit:&[$0.021677$]&[$0.021615$]&[$0.021789$]&\\
\hline
$ \boldsymbol{\Omega_\mathrm{c} h^2}  $ & $0.1199\pm 0.0021$ & $0.1181\pm 0.0012$ & $0.1181\pm 0.0012$ &$  [0.001\,,\,0.1]$\\ 
Bestfit:&[$0.121080$]&[$0.117962$]&[$0.117344$]&\\
\hline
$ \boldsymbol{100\theta_\mathrm{MC}}  $ & $1.04200\pm 0.00068$ & $1.04222\pm 0.00063$ & $1.04220\pm 0.00061
$ &$  [0.5\,,\,10]$ \\ 
Bestfit:&[$1.041856$]&[$1.042178$]&[$1.042044$]&\\
\hline
$ \boldsymbol{\tau_\mathrm{reio}}  $ & $0.064\pm 0.013$ & $0.074\pm 0.010$ & $0.074\pm 0.010$ &$ \mathcal{G}(0.065\,,\,0.015)$\\
Bestfit:&[$0.056704$]&[$0.083767$]&[$0.079482$]&\\
\hline
$ \boldsymbol{\log(10^{10} A_\mathrm{s})}  $ & $3.069\pm 0.025$ & $3.088\pm 0.019$ & $3.088\pm 0.020$ &$  [1.61\,,\,3.91]$\\ 
Bestfit:&[$3.049062$]&[$3.103518$]&[$3.093932$]&\\
\hline
$\boldsymbol{N}$ & $> 82.9$ & $> 85.9$ & $> 87.3$ &$  [10\,,\,200]$\\
Bestfit:&[$198.365310$]&[$199.058100$]&[$179.855190$]&\\
\hline
$ \boldsymbol{\alpha}  $ & -- & $> 0.818$ & $> 1.04$ &$  [0\,,\,7/3]$\\ 
Bestfit:&[$1.673989$]&[$1.934750$]&[$1.767304$]&\\
\hline
$ \boldsymbol{n_\mathrm{s}}  $ & $0.9852^{+0.0051}_{-0.0022}$ & $0.9854^{+0.0048}_{-0.0021}$ & $0.9856^{+0.0047}_{-0.0020}$ \\ 
Bestfit:&[$0.989918$]&[$0.989953$]&[$0.988880$]&\\
\hline
$ \boldsymbol{r}  $ & $0.00092^{+0.00034}_{-0.00083}$ & $0.00113^{+0.00033}_{-0.00084}$ & $ 0.00116^{+0.00032}_{-0.00081}$ \\ 
Bestfit:&[$0.000511$]&[$0.000586$]&[$0.000656$]&\\
\hline
$ \boldsymbol{w_0}  $ & $-0.851^{+0.047}_{-0.15}$ & $-0.924^{+0.013}_{-0.031}$ & $-0.931^{+0.010}_{-0.021}$ \\ 
Bestfit:&[$-0.933625$]&[$-0.942571$]&[$-0.937130$]&\\
\hline
$ \boldsymbol{w_a}  $ & $-0.224^{+0.22}_{-0.071}$ & $-0.114^{+0.047}_{-0.019}$ & $ -0.103^{+0.032}_{-0.015}$ \\ 
Bestfit:&[$-0.099563$]&[$-0.086144$]&[$-0.094306$]&\\
\hline
$ \boldsymbol{\Omega_m} $ & $0.3408^{+0.0075}_{-0.030}$ & $0.3165^{+0.0070}_{-0.0084}$ & $0.3153\pm 0.0065$ \\ 
Bestfit:&[$0.333162$]&[$0.313697$]&[$0.310758$]&\\
\hline
$ \boldsymbol{H_0 } $ & $64.8^{+2.4}_{-0.97}$ & $ 66.66^{+0.70}_{-0.55}$ & $66.78\pm 0.51$ \\
Bestfit:&[$65.606992$]&[$66.857695$]&[$67.066745$]&\\
\hline
$ \boldsymbol{\sigma_8}  $ & $0.808^{+0.022}_{-0.0088}$ & $0.8203^{+0.0085}_{-0.0073}$ & $ 0.8213\pm 0.0075$\\ 
Bestfit:&[$0.817592$]&[$0.831355$]&[$0.823040$]&\\
\hline
\hline 
$\Delta \chi^2_{\texttt{ACT-DR4}}$ & $-0.29$ & $+1.22$ & $+1.29$ \\ 
$\Delta \chi^2_{\texttt{ACT-DR6}}$ & $-0.17$ & $-0.16$ & $-1.00$ \\ 
$\Delta \chi^2_{\texttt{BK18}}$ & $+0.28$ & $-0.34$ & $-0.64$\\
$\Delta \chi^2_{\texttt{BAO}}$ & -- & $-0.05$ & $-0.34$ \\
$\Delta \chi^2_{\texttt{SN}}$ & -- & -- & $-2.26$ \\ 
\hline 
$\Delta \chi^2_{\texttt{tot}}$ & $-0.18$ & $+0.67$ & $-2.95$ \\ 
\hline \hline
\end{tabular} }
\end{center}
\caption{Results for different combinations of data involving  ACT(DR4+DR6), BK18, BAO, and SN. The constraints on parameters are given at 68\% CL, while the upper/lower bounds are given at 95\% CL. For each parameter, the best-fit value is provided as well. Parameters are varied in the prior range (or fixed to the values) reported in the last column of the table. We provide the $\Delta \chi^2$ between the best fit for $\alpha$-attractor quintessential inflation and $\Lambda$CDM for each likelihood used in the analysis.}
\label{tab.results.ACT.alpha_73}
\end{table*}

Following the same narrative thread as the Planck results section, we take an intermediate step and restrict the prior for $\alpha$ to the interval $\alpha\in[0,7/3]$; a choice inspired by Supergravity. The results are presented in \autoref{fig:ACT_alpha_73} and \autoref{tab.results.ACT.alpha_73}. Focusing solely on CMB measurements, from ACT+BK18, we consistently obtain a lower limit $N>82.9$ for the number of e-folds. Once again, this reflects the ACT preference for a larger scalar spectral index constrained to $n_s=0.9852^{+0.0051}_{-0.0022}$. Notice that large values of the number of e-folds lead to small values of the tensor amplitude, and limiting the prior of $\alpha$ does not allow us to balance this effect. Consequently, the value we obtain for $r=0.00092^{+0.00034}_{-0.00084}$ is an order of magnitude lower than the one predicted when $\alpha$ is allowed to vary over a broader range. Additionally, the constraints we obtain for $w_0=-0.851^{+0.047}_{-0.15}$ and $w_a=-0.224^{+0.22}_{-0.071}$ now deviate further from their standard values, deeper into the quintessence regime, albeit with large uncertainties stemming from the fact that $\alpha$ remains unbounded for this combination of data. As discussed at various points previously, the preference for a quintessential DE equation of state produces a constraint $H_0=64.8^{+2.4}_{-0.97}$ km/s/Mpc that is shifted towards smaller values than those characteristic of the standard cosmological model. That being said, the poor constraining power we have on $\alpha$ and the geometrical degeneracy with $\Omega_m$ substantially increase the error-bars on $H_0$.

Including low-redshift data, as usual, breaks the degeneracy between parameters, making the results more precise. For instance, we can now derive an upper limit $\alpha > 0.818$ ($\alpha > 1.04$) for ACT+BK18+BAO(+SN). As a known trend emerging from the different cases considered, including BAO and SN also shifts the constraints for the DE sector towards the cosmological constant. Considering ACT+BK18+BAO and ACT+BK18+BAO+SN, we obtain $w_0 = -0.924^{+0.013}_{-0.031}$, $w_a = -0.114^{+0.047}_{-0.019}$, and $w_0 = -0.931^{+0.010}_{-0.021}$, $w_a = -0.103^{+0.032}_{-0.015}$, respectively. This is also reflected in the results on $H_0$: looking at \autoref{fig:ACT_alpha_73}, we can observe that including low-redshift data pushes the central value of the distribution towards the standard cosmological model. However, reducing the prior on $\alpha$ does not allow us to reach the values preferred in $\Lambda$CDM, forcing instead to smaller $H_0$. Given the reduced uncertainties, this exacerbates the Hubble tension, similar to the case when analyzing Planck data for the same model.

As a final remark, we conclude this section by comparing the best-fit $\chi^2$ of $\alpha$-attractor quintessential inflation with the one obtained within the standard cosmological scenario. The improvement $\Delta\chi^2 \approx -2$ observed for ACT+BK18+BAO+SN persists, even when restricting $\alpha$ to the range $\alpha \in [0, 7/3]$ while the differences in $\Delta\chi^2$ remain marginal for both ACT+BK18 and ACT+BK18+BAO.

\subsection{$\alpha=1$}

\begin{figure*}
    \centering
    \includegraphics[width=1\linewidth]{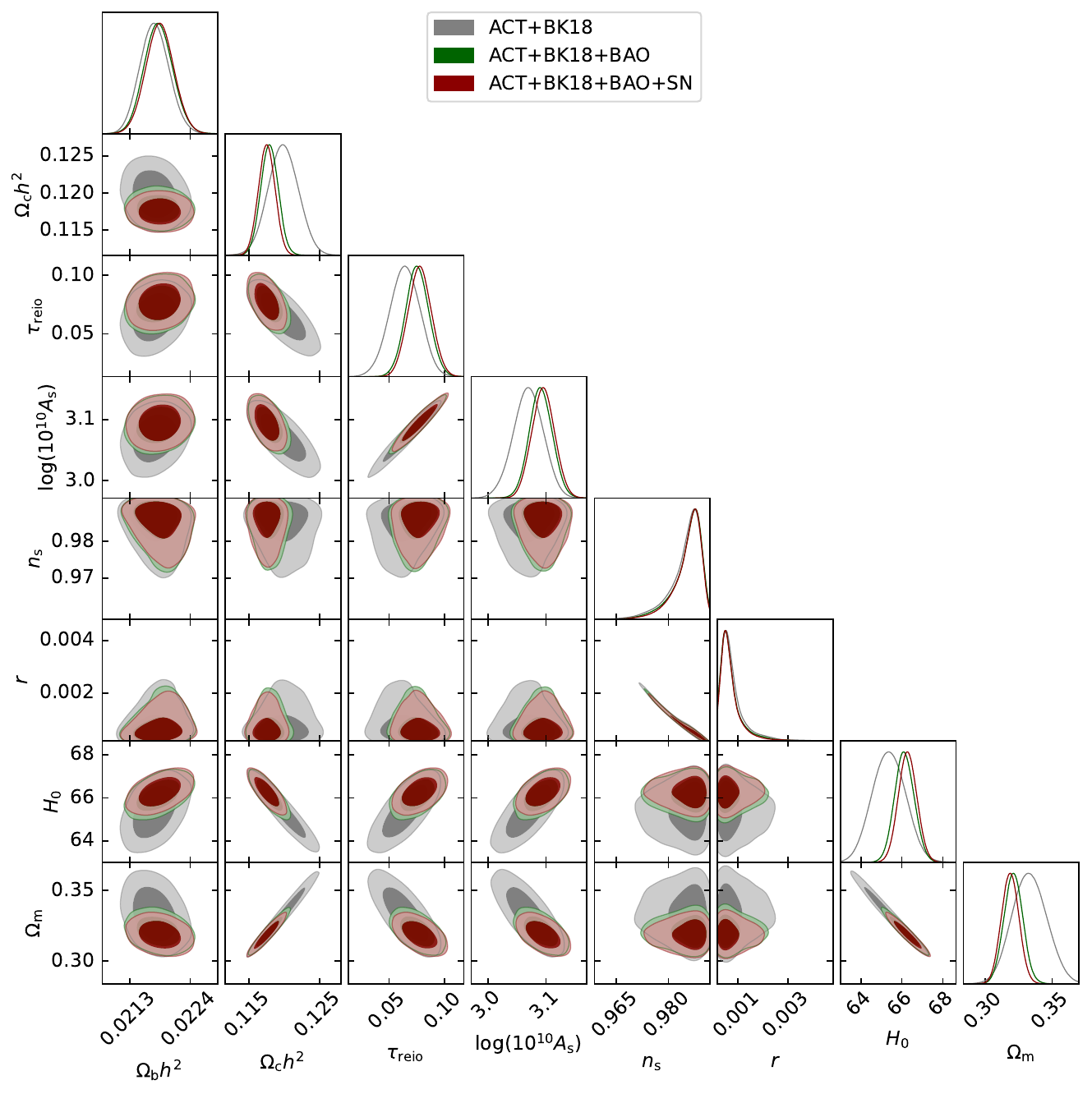}
    \caption{One-dimensional posterior probability distribution functions and two-dimensional contours at 68\% and 95\% CL for the parameters of interest in $\alpha$-attractor quintessential inflation. We consider three reference datasets indicated in the legend fixing $\alpha=1$ (and so $w_0\simeq -0.89$, $w_a=-0.16$ by Eq.~\eqref{eq.w0} and Eq.~\eqref{eq.wa}).}
    \label{fig:ACT_alpha_1}
\end{figure*}

\begin{table*}[htpb!]
\begin{center}
\renewcommand{\arraystretch}{1.5}
\resizebox{0.85 \textwidth}{!}{
\begin{tabular}{l c c c l c c c c c c c c c c c }
\hline
\textbf{Parameter} & \textbf{ ACT+BK18 } & \textbf{ ACT+BK18+BAO } & \textbf{ ACT+BK18+BAO+SN } & \textbf{Prior/Value} \\ 
\hline\hline

$ \boldsymbol{\Omega_\mathrm{b} h^2}  $ & $0.02175\pm 0.00026$ & $0.02182\pm 0.00026$ & $0.02185\pm 0.00025$ &$  [0.005\,,\,0.1]$ \\ 
Bestfit:&[$0.021644$]&[$0.021857$]&[$0.021895$]&\\
\hline
$ \boldsymbol{\Omega_\mathrm{c} h^2}  $ & $0.1199\pm 0.0021$ & $0.1179\pm 0.0012$ & $0.1175\pm 0.0012$ &$  [0.001\,,\,0.1]$\\ 
Bestfit:&[$0.119138$]&[$0.116887$]&[$0.117471$]&\\
\hline
$ \boldsymbol{100\theta_\mathrm{MC}}  $ & $1.04201\pm 0.00067$ & $1.04225\pm 0.00062$ & $1.04228\pm 0.00062$ &$  [0.5\,,\,10]$ \\ 
Bestfit:&[$1.042161$]&[$1.042115$]&[$1.042554$]&\\
\hline
$ \boldsymbol{\tau_\mathrm{reio}}  $ & $0.064\pm 0.013$ & $0.075\pm 0.010$ & $0.078\pm 0.010$  &$ \mathcal{G}(0.065\,,\,0.015)$\\ 
Bestfit:&[$0.071574$]&[$0.088287$]&[$0.083763$]&\\
\hline
$ \boldsymbol{\log(10^{10} A_\mathrm{s})}  $ & $3.069\pm 0.026$ & $3.091\pm 0.020$ & $3.095\pm 0.019$ &$  [1.61\,,\,3.91]$\\ 
Bestfit:&[$3.080345$]&[$3.115583$]&[$3.101780$]&\\
\hline
$\boldsymbol{N}$ & $>81.8$ & $> 86.1$ & $ > 89.1$ &$  [10\,,\,200]$\\ 
Bestfit:&[$191.694030$]&[$131.671060$]&[$174.979570$]&\\
\hline
$ \boldsymbol{\alpha}  $ & -- & --  & --  &$ \alpha=1$\\
\hline
$ \boldsymbol{n_\mathrm{s}}  $ & $0.9852^{+0.0051}_{-0.0022}$ & $0.9856^{+0.0048}_{-0.0020}$ & $0.9857^{+0.0045}_{-0.0020}$\\ 
Bestfit:&[$0.989567$]&[$0.984811$]&[$0.988570$]&\\
\hline
$ \boldsymbol{r}  $ & $0.00073^{+0.00019}_{-0.00051}$ & $0.00068^{+0.00017}_{-0.00046}$ & $0.00067^{+0.00016}_{-0.00043}$\\ 
Bestfit:&[$0.000327$]&[$0.000692$]&[$0.000392$]&\\
\hline
$ \boldsymbol{w_0}  $ & -- & -- & -- & $w_0\simeq-0.89$ \\ 
\hline
$ \boldsymbol{w_a}  $ & -- & -- & -- & $w_a\simeq-0.16$ \\ 
\hline
$ \boldsymbol{\Omega_m} $ & $0.333\pm 0.013$ & $0.3211\pm 0.0069$ & $0.3187\pm 0.0064$ \\ 
Bestfit:&[$0.329088$]&[$0.315778$]&[$0.317430$]&\\
\hline
$ \boldsymbol{H_0 } $ & $65.37\pm 0.80$ & $66.13\pm 0.48$ & $66.28\pm 0.45$ \\ 
Bestfit:&[$65.555668$]&[$66.438988$]&[$66.413532$]&\\
\hline
$ \boldsymbol{\sigma_8}  $ & $0.8129\pm 0.0072$ & $0.8150\pm 0.0070$ & $0.8151\pm 0.0070$ \\ 
Bestfit:&[$0.816935$]&[$0.820810$]&[$0.818648$]&\\

\hline 
$\Delta \chi^2_{\texttt{ACT-DR4}}$ & $+0.28$ & $+2.68$ & $+2.02$ \\ 
$\Delta \chi^2_{\texttt{ACT-DR6}}$ & $-0.72$ & $-0.68$ & $-1.15$ \\
$\Delta \chi^2_{\texttt{BK18}}$ & $+1.08$ & $-0.53$ & $-0.52$\\
$\Delta \chi^2_{\texttt{BAO}}$ & -- & $+0.34$ & $+0.14$ \\ 
$\Delta \chi^2_{\texttt{SN}}$ & -- & -- & $-1.57$ \\ 
\hline 
$\Delta \chi^2_{\texttt{tot}}$ & $+0.64$ & $+1.81$ & $ -1.09$ \\ 
\hline \hline
\end{tabular} }
\end{center}
\caption{Results for different combinations of data involving  ACT(DR4+DR6), BK18, BAO, and SN. The constraints on parameters are given at 68\% CL, while the upper/lower bounds are given at 95\% CL. For each parameter, the best-fit value is provided as well. Parameters are varied in the prior range (or fixed to the values) reported in the last column of the table. We provide the $\Delta \chi^2$ between the best fit for $\alpha$-attractor quintessential inflation and $\Lambda$CDM for each likelihood used in the analysis.}
\label{tab.results.ACT.alpha_1}
\end{table*}

We conclude our investigation considering, as a final case, a model inspired by Starobinsky and Higgs inflation, where $\alpha=1$. As explained in previous sections, this choice constrains the DE equation of state in the quintessence regime, setting $w_0=-0.89$ and $w_a=-0.16$. Our findings are summarized in \autoref{tab.results.ACT.alpha_1} and depicted in the contour plot \autoref{fig:ACT_alpha_1}. The results do not significantly deviate from expectations based on the discussions in the preceding sections. Specifically, focusing on temperature and polarization spectra measurements, as well as lensing data from ACT and BK18, the overall trend towards somewhat large values of $n_s$ is mostly confirmed, leading to a lower limit of $N>81.8$. Furthermore, fixing $\alpha$ reduces the amplitude of gravitational waves predicted in this model. For this particular dataset, it is constrained within the interval $r=0.00073^{+0.00019}_{-0.00051}$. Similarly, restricting ourselves to quintessential values for $w_0$ and $w_a$ influences the late-time evolution of the Universe, favoring a lower value of $H_0=65.37\pm0.80$ and the well-known degeneracy between $H_0$ and $\Omega_m$.

As consistently observed in this study, BAO and SN contribute to precisely determining $\Omega_m$, breaking the $\Omega_m\, h^2$ degeneracy typical of CMB measurements and allowing for more precise constraints on parameters. That said, the constraints on $H_0=66.13\pm0.48$ km/s/Mpc and $H_0=66.28\pm0.45$ km/s/Mpc, obtained with ACT+BK18+BAO and ACT+BK18+BAO+SN, respectively, clearly show that fixing $\alpha$ (and consequently $w_0$ and $w_a$) prevents us from recovering values of $H_0\sim67.5$ km/s/Mpc typical in $\Lambda$CDM, thereby exacerbating the Hubble tension. We emphasize once again that the reason for this lies in fixing a quintessential component for DE, which alters the late evolution of the Universe, potentially impacting the value we can obtain for the angular diameter distance from the CMB, $D_A(z_{\rm CMB})$. This distance is determined by the simple trigonometric relation $D_A(z_{\rm CMB}) = r_s(z_{\rm CMB}) / \theta_{s}$, where $\theta_{s}$ is the angular scale corresponding to the position of the first acoustic peak in the temperature anisotropy CMB spectrum (well-measured by CMB experiments), and $r_s(z_{\rm CMB})$ represents the value of the sound horizon, which is sensitive to physics prior to recombination that, in our model, remains mostly the same as in $\Lambda$CDM. Therefore, both $r_s(z_{\rm CMB})$ and $\theta_{s}(z_{\rm CMB})$ are essentially fixed, so that $D_A(z_{\rm CMB})$ cannot change as well. As a result, to compensate for a quintessential DE component, we need to adjust other parameters involved in the calculation of $D_A(z_{\rm CMB})$, namely $H_0$ and $\Omega_m$. As both BAO and SN measurements accurately determine $\Omega_m$, this parameter cannot shift significantly. Hence, we are left with only one option: in the presence of a quintessence DE component, when considering CMB and low-redshift data together, to maintain a constant $D_A(z_{\rm CMB})$ and preserve $\theta_{s}$, we need to decrease the value of $H_0$. This is what we observe here as well as in the analysis of Planck data.

Regarding the $\chi^2$ analysis, the results presented in \autoref{tab.results.ACT.alpha_1} align with the conclusions drawn in the previous two cases, confirming an overall improvement in the fit to ACT+BK18+BAO+SN (now $\Delta\chi^2\sim -1$).

\begin{table}
    \centering
    \renewcommand{\arraystretch}{1.5}
    \resizebox{\columnwidth}{!}{\begin{tabular}{lc|c}
        \hline
       \textbf{Prior}  & \textbf{P18+BK18+BAO+SN} & \textbf{ACT+BK18+BAO+SN} \\
        &  $\Delta \chi^2_{\rm tot} $& $ \Delta \chi^2_{\rm tot} $\\
        \hline \hline
       $\alpha\in[0,\alpha_{\rm max}\gg1]$  & $-0.06$ & $-2.11$ \\
       $\alpha\in[0,7/3]$  & $-0.20$ & $-2.95$ \\
       $\alpha=1$  & $+3.83$ & $-1.09$ \\
       \hline \hline
    \end{tabular}}
    \caption{Table summarizing $\Delta\chi^2_{\rm tot}$ for the three $\alpha$ prior cases and the two CMB experiments in combination with BAO and SN data.}
    \label{tab.summary}
\end{table}

Overall, when considering the same three cases in light of the ACT small-scale CMB data, BK18 likelihood, and low-redshift measurements, we can conclude that the ACT preference toward a scale-invariant spectrum persists. This preference significantly shifts the results obtained for the inflationary parameters. A large $n_s$ requires $N$ to approach very large values. Depending on the prior adopted for $\alpha$, this can influence the predictions for the tensor amplitude as well as alter the constraints on late-time cosmology. Restricting the prior to models inspired by supergravity or fixing $\alpha=1$ following Starobinsky or Higgs inflation leads to very small values of $r$, while simultaneously pushing the model into a DE quintessence regime lowering $H_0$. Regardless of the prior adopted for $\alpha$, an overall improvement in the fit to ACT+BK18+BAO+SN is observed compared to $\Lambda$CDM. However, models that allow greater freedom in the value of this parameter appear to be the only ones able to not exacerbate the Hubble tension significantly.

\section{Conclusions}
\label{sec:concl}
In this study, we focused on $\alpha$-attractor quintessential inflation. From a theoretical standpoint, the model has desirable features such as robustness against quantum corrections, shift symmetry, and attractor behavior that ensures insensitivity to initial conditions. On the observational front, the universality and attractor properties of $\alpha$-attractor quintessential inflation establish direct relationships between inflationary observables (like the scalar tilt $n_s$ and the tensor-to-scalar ratio $r$) and the parameters describing late-time DE equation of state, $w_0$ and $w_a$. Consequently, predictions for the inflationary epoch of the theory influence the dynamics of the Universe at late times, predicting a dynamic DE component lying in the quintessential regime. These predictions have repercussions on various cosmological observables, both at early and late times. For instance, the spectrum of primordial scalar and tensor perturbations produced during inflation can lead to signatures in the cosmic microwave background radiation, while the quintessential behavior of the DE at late times influences local distances in the Universe. 

The central question we aimed to address in this paper is whether this model aligns with the most recent CMB measurements as well as with low-redshift data such as BAO and SN. To answer this question, we implement the theoretical predictions of the model in the Boltzmann solver code \texttt{CAMB} and perform a full MCMC analysis, which assumes this model from the outset. To the best of our knowledge, this work represents one of the first studies where quintessential inflation is assumed from the starting point in the analysis, making it a pivotal analysis from which we derived some interesting conclusions.

\subsection{Conclusions based on Planck CMB measurements}

In \autoref{sec:planck}, we examined the model in light of temperature polarization and lensing measurements released by the Planck satellite, as well as B-mode polarization data from the BICEP/Keck collaboration. Additionally, we considered local Universe probes such as BAO and SN measurements for the PantheonPlus catalog. We analyzed these data in the context of three theoretical scenarios, differing in the priors adopted for the parameter $\alpha$. Initially, we varied $\alpha$ continuously in the range $\alpha\in[0, \alpha_{\rm max}\gg1]$, and subsequently constrained it to a continuous range $\alpha=[0, 7/3]$. This choice was inspired by models of supergravity, where $\alpha$ is related to Poincaré disks of the symmetry group and takes on discrete values up to $\alpha_{\rm max}=7/3$. Finally, we considered the case corresponding to the Starobinsky inflation setting $\alpha=1$. The most relevant conclusions drawn from the analysis of Planck (and low-redshift) data for the three cases are the following:

\begin{itemize}

\item When adopting a broad prior $\alpha_{\rm max}\gg1$, the data show good agreement with the model. We consistently observe an improvement in the $\chi^2$ of the best fit when compared to the $\chi^2$ of the standard cosmological model for the same combination of data. Although such improvement is not sufficient to conclude that $\alpha$-attractor quintessential inflation is favored over $\Lambda$CDM, these results demonstrate that connecting the two epochs of accelerated expansion in the Universe within a framework that, in terms of assumptions, is certainly more economical than the standard cosmological model, does not lead to a deterioration of the fit to cosmological observables, neither at early times (i.e., CMB) nor at late times (i.e., BAO and SN), see also \autoref{tab.summary}. Given that in this model, we are constrained to a regime of quintessential DE (well-known for predicting values of $H_0$ that are typically smaller than those observed in the standard cosmological model), it is important to note that in this case, we recover values of $H_0$ consistent with $\Lambda$CDM, without exacerbating (or reducing) the well-known tension surrounding the value of this parameter.

\item When focusing on models inspired by supergravity and restricting the prior to $\alpha\in[0,7/3]$, the situation becomes somewhat more intricate. On the one hand, we still observe an improvement in the fit compared to $\Lambda$CDM for most data combinations, see \autoref{tab.summary}. However, limiting $\alpha$ forces us into a more quintessential region of the parameter space, reducing our freedom to move toward the cosmological constant regime (see also Eq.~\eqref{eq.w0} and Eq.~\eqref{eq.wa}). This also implies that the values inferred for $H_0$ in this case are lower than in the standard cosmological model, increasing the Hubble tension.

\item The situation worsens when fixing $\alpha=1$. In this case, the model fixes DE parameters to non-standard values $w_0=-0.89$ and $w_a=-0.16$ with no freedom left. Since in our model the value of the sound horizon at recombination remains the same as in $\Lambda$CDM, to compensate for the effects produced by a quintessential DE component in the angular diameter distance from the CMB, $D_A(z_{\rm CMB})$, and preserve $\theta_{s}=r_s(z_{\rm CMB}) / D_A(z_{\rm CMB})$, we need to adjust other parameters involved in the calculation of $D_A(z_{\rm CMB})$, namely $H_0$ and $\Omega_m$. As both BAO and SN measurements accurately determine $\Omega_m$, when considering CMB and low-redshift data together, to maintain a constant $D_A(z_{\rm CMB})$ and preserve $\theta_{s}$, we need to decrease the value of $H_0$, significantly increasing the Hubble tension. Additionally, in the presence of such a quintessential component, we observe a significant worsening in the fit compared to $\Lambda$CDM when including late-time data as evident from \autoref{tab.summary}.

\end{itemize}

Based on these considerations, we can certainly conclude that a quintessential model allowing more freedom in $\alpha$ seems to be the preferred choice based on current observations. Interestingly, when $\alpha > 1$ the model predicts values of the tensor amplitude $r$ that could be visible by future CMB probes.

\subsection{Conclusions based on ACT CMB measurements}

In \autoref{sec:ACT}, we explored the same three scenarios of quintessential inflation, replacing the Planck data with temperature, polarization, and lensing measurements from ACT. This independent experiment provides a significant opportunity to test our results without relying exclusively on Planck. Additionally, ACT exhibits a preference toward a scale-invariant Harrison-Zel'dovich spectrum with $n_s \approx 1$, enabling us to investigate the implications of approaching scale invariance. This preference substantially alters the results obtained for inflationary parameters, as achieving $n_s \approx 1$ pushes $N$ to very large values. Depending on the prior adopted for $\alpha$, this can influence both predictions for the tensor amplitude and the constraints on late-time cosmology:

\begin{itemize}

\item Exceedingly large values $\alpha \gg 1$ were ruled out for Planck as they increased the tensor amplitude $r$ too much. For ACT, large values of $\alpha$ can always be balanced by large values of $N$ that shift $n_s$ towards a scale-invariant spectrum. Such values are strongly favored by ACT data, leading to a clear degeneracy between these two parameters. Additionally, values $\alpha \gg 1$ drive the parameters describing the DE equation of state towards the cosmological constant regime, providing a good fit to the late-time Universe, as well. Therefore, we are now free to move towards $\alpha \gg 1$ without compromising the result on the tensor amplitude, which remains consistent with B-mode polarization measurements.

\item Restricting the prior to models inspired by supergravity reduces our freedom to vary the parameter $\alpha$. Consequently, we can no longer compensate for large values of $N$, significantly reducing the amplitude of tensor perturbations. Furthermore, akin to Planck, at late times we are constrained to a quintessential DE regime. For this reason, the model prefers lower values of $H_0$ compared to $\Lambda$CDM.

\item Both of these issues are exacerbated when fixing $\alpha=1$. This further reduces the amplitude of gravitational waves (now just $r=12/N^2$) while leaving us with no freedom in late-time parameters. With the model now grounded in the quintessential regime, we infer significantly smaller values of $H_0$ compared to $\Lambda$CDM, thereby increasing the Hubble tension.

\end{itemize}

Regardless of the prior adopted for $\alpha$, an overall improvement in the fit to ACT+BK18+BAO+SN is observed compared to $\Lambda$CDM, see also \autoref{tab.summary}. This improvement is mostly coming from ACT-DR6 lensing data and SN measurements. However, the conclusion reached for the Planck satellite remains mostly true: models that allow greater freedom in $\alpha$ seem to be the preferred choice to avoid further exacerbating the Hubble tension.

\begin{acknowledgments}
EDV is supported by a Royal Society Dorothy Hodgkin Research Fellowship. EL is supported in part by the U.S.\ Department of Energy, Office of Science, Office of High Energy Physics, under contract no.\ DE-AC02-05CH11231. 
This article is based upon work from COST Action CA21136 Addressing observational tensions in cosmology with systematics and fundamental physics (CosmoVerse) supported by COST (European Cooperation in Science and Technology). We acknowledge IT Services at The University of Sheffield for the provision of services for High Performance Computing. 
\end{acknowledgments}

\clearpage
\appendix
\widetext

\section*{Supplementary Material}
\label{sec:supmat}
\vspace{0.5 cm}
\subsection{Results for $\Lambda$CDM}

\begin{table*}[htpb!]
\begin{center}
\renewcommand{\arraystretch}{1.5}
\resizebox{1 \textwidth}{!}{
\begin{tabular}{l | c c c | c c c c c c c c c c c c }
\hline
\textbf{Parameter} & \textbf{ P18+BK18 } & \textbf{ P18+BK18+BAO } & \textbf{ P18+BK18+BAO+SN }  & \textbf{ ACT+BK18 } & \textbf{ ACT+BK18+BAO } & \textbf{ ACT+BK18+BAO+SN } \\
\hline\hline

$ \boldsymbol{\Omega_\mathrm{b} h^2}$ & $0.02236\pm 0.00015$ & $0.02245\pm 0.00013$ & $0.02242\pm 0.00013$ & $0.02160\pm 0.00030$ & $0.02166\pm 0.00029$ & $0.02161\pm 0.00030$ \\ 
Bestfit:&[$0.022297$]&[$0.022430$]&[$0.022360$]& [$0.021677$]&[$0.021579$]&[$0.021363$]\\
\hline
$ \boldsymbol{\Omega_\mathrm{c} h^2}  $ & $0.1202\pm 0.0012$ & $0.11908\pm 0.00090$ & $0.11948\pm 0.00087$ & $0.1195\pm 0.0021$ & $0.1186\pm 0.0012$ & $0.1192\pm 0.0012$ \\
Bestfit:&[$0.120577$]&[$0.119873$]&[$0.120335$]& [$0.117693$]&[$0.117486$]&[$0.118969$]\\
\hline
$ \boldsymbol{100\theta_\mathrm{MC}}  $ & $1.04089\pm 0.00030$ & $1.04104\pm 0.00029$ & $1.04099\pm 0.00029$ & $1.04207\pm 0.00066$ & $1.04217\pm 0.00062$ & $1.04212\pm 0.00064$ \\ 
Bestfit:&[$1.041018$]&[$1.040919$]&[$1.041159$]& [$1.041950$]&[$1.041980$]&[$1.042078$]\\
\hline
$ \boldsymbol{\tau_\mathrm{reio}}  $ & $0.0547\pm 0.0075$ & $0.0593\pm 0.0073$ & $0.0578\pm 0.0071$ & $0.068\pm 0.014$ & $0.073\pm 0.011$ & $ 0.071\pm 0.011$ \\
Bestfit:&[$0.051398$]&[$0.055330$]&[$0.053551$]& [$0.087335$]&[$0.088131$]&[$0.078100$]\\
\hline
$ \boldsymbol{\log(10^{10} A_\mathrm{s})}  $ & $3.046\pm 0.014$ & $3.054\pm 0.014$ & $3.051\pm 0.014$ & $3.068\pm 0.025$ & $3.078\pm 0.019$ & $3.074\pm 0.019$ \\ 
Bestfit:&[$3.036944$]&[$3.050957$]&[$3.059518$]& [$3.092919$]&[$3.104948$]&[$3.083902$]\\
\hline
$ \boldsymbol{n_\mathrm{s}}  $ & $0.9647\pm 0.0041$ & $0.9675\pm 0.0037$ & $0.9666\pm 0.0036$ & $0.996\pm 0.012$ & $0.997\pm 0.011$ & $0.997\pm 0.011$ \\ 
Bestfit:&[$0.962858$]&[$0.968359$]&[$0.965014$]& [$1.009197$]&[$1.000990$]&[$1.001662$]\\
\hline
$ \boldsymbol{\Omega_m} $ & $0.3164\pm 0.0074$ & $0.3095\pm 0.0054$ & $0.3119\pm 0.0053$ & $0.314\pm 0.013$ & $0.3076\pm 0.0067$ & $0.3112\pm 0.0065$ \\
Bestfit:&[$0.31843521$]&[$0.31420408$]&[$0.31630986$]& [$0.30338967$]&[$0.3026908$]&[$0.31157462$]\\
\hline
$ \boldsymbol{H_0 } $ & $67.29\pm 0.53$ & $67.78\pm 0.41$ & $67.61\pm 0.39$ & $67.26\pm 0.89$ & $67.68\pm 0.50$ & $67.42\pm 0.48$ \\
Bestfit:&[$67.134152$]&[$67.450285$]&[$67.317361$]& [$67.933919$]&[$67.938103$]&[$67.265448$]\\
\hline
$ \boldsymbol{\sigma_8}  $ & $0.8124\pm 0.0058$ & $0.8124\pm 0.0058$ & $0.8126\pm 0.0058$ & $0.8335\pm 0.0080$ & $0.8345\pm 0.0078$ & $0.8349\pm 0.0081$ \\ 
Bestfit:&[$0.809788$]&[$0.814193$]&[$0.818711$]& [$0.841342$]&[$0.843470$]&[$0.841070$]\\

\hline 
$\Delta \chi^2_{\texttt{low-TT}}$ & $23.73$ & $22.63$ & $23.24$ & -- & -- & -- \\  
$\Delta \chi^2_{\texttt{low-EE}}$ & $395.79$ & $396.23$ & $395.99$ & --& -- & -- \\  
$\Delta \chi^2_{\texttt{TTTEEE}}$ & $2346.21$ & $2345.13$ & $2344.82$ & -- & -- & -- \\  
$\Delta \chi^2_{\texttt{lensing}}$ & $8.87$ & $8.90$ & $9.25$ & -- & -- & -- \\  
$\chi^2_{\texttt{BK18}}$ & $537.45$ & $538.05$ & $536.92$ & $535.62$ & $536.39$ & $536.63$ \\ 
$\chi^2_{\texttt{BAO}}$ & -- & $23.52$ & $23.93$ & -- & $20.59$ & $21.06$ \\ 
$\chi^2_{\texttt{SN}}$ & -- & -- & $1410.80$ & -- & -- & $1411.62$ \\ 
$\chi^2_{\texttt{ACT-DR4}}$ & -- & -- & -- & $281.75$ & $281.41$ & $281.29$ \\  
$\chi^2_{\texttt{ACT-DR6}}$ & -- & -- & -- & $15.29$ & $14.63$ & $15.22$ \\ 
\hline 
$\chi^2_{\texttt{tot}}$ & $3312.05$ & $3334.46$ & $4744.95$ & $832.66$ & $853.02$ & $2265.82$ \\ 
\hline \hline
\end{tabular} }
\end{center}
\caption{Results for different combinations of data involving Planck-2018, ACT(DR4+DR6), BK18, BAO, and SN for the $\Lambda$CDM model. The constraints on parameters are given at 68\% CL, while the upper/lower bounds are given at 95\% CL. For each parameter, the best-fit value is provided as well. 
We provide the best-fit $\chi^2$ for each likelihood used in the analysis.}
\label{ tab.results.LCDM }
\end{table*}

\clearpage

\subsection{Parameter Best-fit Comparison}

\begin{table*}[htpb!]
\begin{center}
\renewcommand{\arraystretch}{1.5}
\resizebox{1 \textwidth}{!}{
\begin{tabular}{l | c c c | c c c c c c c c c c c c }
\hline
\textbf{Parameter} & \textbf{ P18+BK18 } & \textbf{ P18+BK18+BAO } & \textbf{ P18+BK18+BAO+SN }  & \textbf{ ACT+BK18 } & \textbf{ ACT+BK18+BAO } & \textbf{ ACT+BK18+BAO+SN } \\
\hline\hline

$ \boldsymbol{\Omega_\mathrm{b} h^2}$: & & & & & & \\
Bestfit $\Lambda$CDM:& $0.022297$ & $0.022430$ & $0.022360$ &  $0.021677$ & $0.021579$ & $0.021363$ \\
Bestfit $\alpha$ free &$0.022380$&$0.022534$&$0.022446$ &$0.021748$&$0.021825$&$0.021877$ \\
Bestfit $\alpha\in[0\,,\,7/3]$ &$0.022478$&$0.022555$&$0.022412$ &$0.021677$&$0.021615$&$0.021789$\\
Bestfit $\alpha=1$ &$0.022504$&$0.022418$&$0.022494$&$0.021644$&$0.021857$&$0.021895$\\

\hline

$ \boldsymbol{\Omega_\mathrm{c} h^2}  $ & & & & & & \\
Bestfit $\Lambda$CDM: & $0.120577$ & $0.119873$ & $0.120335$ &  $0.117693$ & $0.117486$ & $0.118969$ \\
Bestfit $\alpha$ free &$0.120385$&$0.119258$&$0.119496$&$0.119103$&$0.118734$&$0.117713$\\
Bestfit $\alpha\in[0\,,\,7/3]$&$0.119982$&$0.118382$&$0.118677$ &$0.121080$&$0.117962$&$0.117344$\\
Bestfit $\alpha=1$ &$0.119260$&$0.118064$&$0.118799$&$0.119138$&$0.116887$&$0.117471$\\

\hline

$ \boldsymbol{100\theta_\mathrm{MC}}  $ & & & & & & \\ 
Bestfit $\Lambda$CDM:& $1.041018$ & $1.040919$ & $1.041159$ &  $1.041950$ & $1.041980$ & $1.042078$ \\
Bestfit $\alpha$ free&$1.040969$&$1.040933$&$1.041011$ &$1.042179$&$1.042366$&$1.042224$\\
Bestfit $\alpha\in[0\,,\,7/3]$&$1.040871$&$1.040984$&$1.041175$&$1.041856$&$1.042178$&$1.042044$\\
Bestfit $\alpha=1$&$1.040867$&$1.041104$&$1.041049$&$1.042161$&$1.042115$&$1.042554$\\

\hline

$ \boldsymbol{\tau_\mathrm{reio}}  $ & & & & & & \\
Bestfit $\Lambda$CDM:& $0.051398$ & $0.055330$ & $0.053551$ &  $0.087335$ & $0.088131$  & $0.078100$ \\
Bestfit $\alpha$ free &$0.052943$&$0.060925$&$0.057164$ &$0.069651$&$0.071363$&$0.081290$\\
Bestfit $\alpha\in[0\,,\,7/3]$&$0.057438$&$0.061664$&$0.064106$&$0.056704$&$0.083767$&$0.079482$\\
Bestfit $\alpha=1$ &$0.052012$&$0.059842$&$0.060756$&$0.071574$&$0.088287$&$0.083763$\\

\hline

$ \boldsymbol{\log(10^{10} A_\mathrm{s})} $& & & & & & \\ 
Bestfit $\Lambda$CDM: & $3.036944$ & $3.050957$ & $3.059518$ &  $3.092919$ & $3.104948$ & $3.083902$ \\
Bestfit $\alpha$ free &$3.053140$&$3.064418$&$3.056218$ &$3.074046$&$3.083627$&$3.098476$\\
Bestfit $\alpha\in[0\,,\,7/3]$&$3.056608$&$3.057068$&$3.063715$&$3.049062$&$3.103518$&$3.093932$\\
Bestfit $\alpha=1$&$3.048496$&$3.060888$&$3.052388$&$3.080345$&$3.115583$&$3.101780$\\

\hline

$ \boldsymbol{n_\mathrm{s}}  $ & & & & & & \\
Bestfit $\Lambda$CDM: & $0.962858$ & $0.968359$ & $0.965014$ &  $1.009197$ & $1.000990$ & $1.001662$ \\
Bestfit $\alpha$ free &$0.964861$&$0.970651$&$0.967247$ &$0.989786$&$0.988495$&$0.989571$\\
Bestfit $\alpha\in[0\,,\,7/3]$&$0.967309$&$0.970603$&$0.966853$&$0.989918$&$0.989953$&$0.988880$\\
Bestfit $\alpha=1$&$0.963968$&$0.971059$&$0.968245$&$0.989567$&$0.984811$&$0.988570$\\

\hline \hline
\end{tabular} }
\end{center}
\caption{Comparison between the best fit parameters for different models and data.}
\label{ tab.results.LCDM }
\end{table*}

\clearpage

\bibliographystyle{apsrev4-1}
\bibliography{main.bib}
\end{document}